\documentclass[twocolumn]{aastex631}

\usepackage{bm}
\usepackage{mathtools}
\DeclarePairedDelimiter{\abs}{\lvert}{\rvert}
\shorttitle{Simulated Dwarf Satellites of the Milky Way}
\shortauthors{Hirai et al.}
\graphicspath{{./}{figures/}}

\begin{document}

\title{Chemo-dynamical Evolution of Simulated Satellites for a Milky Way-like Galaxy}

\author[0000-0002-5661-033X]{Yutaka Hirai}
\altaffiliation{JSPS Research Fellow}
\affiliation{Department of Physics and Astronomy, University of Notre Dame, 225 Nieuwland Science Hall, Notre Dame, IN 46556, USA}
\affiliation{Astronomical Institute, Tohoku University, 6-3 Aoba, Aramaki, Aoba-ku, Sendai, Miyagi 980-8578, Japan}
\affiliation{Joint Institute for Nuclear Astrophysics, Center for the Evolution of the Elements (JINA-CEE), USA}

\author[0000-0001-6196-5162]{Evan N. Kirby}
\affiliation{Department of Physics and Astronomy, University of Notre Dame, 225 Nieuwland Science Hall, Notre Dame, IN 46556, USA}

\author[0000-0002-9053-860X]{Masashi Chiba}
\affiliation{Astronomical Institute, Tohoku University, 6-3 Aoba, Aramaki, Aoba-ku, Sendai, Miyagi 980-8578, Japan}

\author[0000-0002-8758-8139]{Kohei Hayashi}
\affiliation{National Institute of Technology, Sendai College, 48 Nodayama, Medeshima-Shiote, Natori, Miyagi 981-1239, Japan}
\affiliation{Astronomical Institute, Tohoku University, 6-3 Aoba, Aramaki, Aoba-ku, Sendai, Miyagi 980-8578, Japan}
\affiliation{Institute for Cosmic Ray Research, The University of Tokyo, 5-1-5 Kashiwa-no-Ha, Kashiwa, Chiba 277-8582, Japan}

\author[0000-0001-5261-4336]{Borja Anguiano}
\affiliation{Department of Physics and Astronomy, University of Notre Dame, 225 Nieuwland Science Hall, Notre Dame, IN 46556, USA}

\author[0000-0001-8226-4592]{Takayuki R. Saitoh}
\affiliation{Department of Planetology, Graduate School of Science, Kobe University, 1-1 Rokkodai-cho, Nada-ku, Kobe, Hyogo 657-8501, Japan}

\author[0000-0003-4656-0241]{Miho N. Ishigaki}
\affiliation{National Astronomical Observatory of Japan, 2-21-1 Osawa, Mitaka, Tokyo 181-8588, Japan}
\affiliation{Kavli Institute for the Physics and Mathematics of the Universe, The University of Tokyo, 5-1-5 Kashiwa-no-Ha, Kashiwa, Chiba 277-8582, Japan}

\author[0000-0003-4573-6233]{Timothy C. Beers}
\affiliation{Department of Physics and Astronomy, University of Notre Dame,
225 Nieuwland Science Hall, Notre Dame, IN 46556, USA}
\affiliation{Joint Institute for Nuclear Astrophysics, Center for the Evolution of the Elements (JINA-CEE), USA}

\begin{abstract}

The chemical abundances of Milky Way's satellites reflect their star formation histories (SFHs), yet, due to the difficulty of determining the ages of old stars, the SFHs of most satellites are poorly measured. Ongoing and upcoming surveys will obtain around ten times more medium-resolution spectra for stars in satellites than are currently available. To correctly extract SFHs from large samples of chemical abundances, the relationship between chemical abundances and SFHs needs to be clarified. Here, we perform a high-resolution cosmological zoom-in simulation of a Milky Way-like galaxy with detailed models of star formation, supernova feedback, and metal diffusion. We quantify SFHs, metallicity distribution functions, and the $\alpha$-element (Mg, Ca, and Si) abundances in satellites of the host galaxy. We find that star formation in most simulated satellites is quenched before infalling to their host. Star formation episodes in simulated satellites are separated by a few hundred Myr owing to supernova feedback; each star formation event produces groups of stars with similar [$\alpha$/Fe] and [Fe/H]\@. We then perform a mock observation of the upcoming Subaru Prime Focus Spectrograph (PFS) observations. We find that Subaru PFS will be able to detect distinct groups of stars in [$\alpha$/Fe] vs. [Fe/H] space, produced by episodic star formation. This result means that episodic SFHs can be estimated from the chemical abundances of $\gtrsim$ 1,000 stars determined with medium-resolution spectroscopy.

\end{abstract}

\section{Introduction}
\label{sec:intro}

Satellite galaxies of the Milky Way (MW) are crucial for understanding galaxy formation \citep[e.g.,][]{Bullock2017}. Many satellites possess ancient stars; the histories of star formation and galaxy assembly are imprinted in the chemo-dynamical properties of such satellites. In our Galaxy, thanks to their relatively close distances from the Sun ($\approx$ 20--200 kpc), we can observe the chemo-dynamical properties of individual stars in the MW's satellites \citep[e.g.,][]{Tolstoy2004, Battaglia2006, Kirby2011b}.

Over 100 dwarf galaxies are identified within 3 Mpc \citep[e.g.,][]{Mcconnachie2012, Simon2019}. Galaxies with stellar masses ($M_{*}$) less than $\approx10^9\,M_{\sun}$ are typically categorized as dwarf galaxies. Among them, gas-free dwarf galaxies with $M_{*} \gtrsim10^5\,M_{\sun}$ are called classical dwarf spheroidal galaxies (dSphs), while those with $M_{*} \lesssim10^5\,M_{\sun}$ are identified as ultrafaint dwarf galaxies (UFDs). Many of the dwarf galaxies in the Local Group are satellites of the MW or M31;  interactions with their more massive hosts could affect the chemo-dynamical properties of these satellites \citep{Genina19,Kvasova24}.

Satellites exhibit a wide variety of star formation histories (SFHs) and chemical abundances. The SFHs of Local Group dwarf galaxies can be derived by color-magnitude diagrams \citep[CMDs, e.g.,][]{deBoer2012a,deBoer2012b,Weisz2014,Ren2024}. \citet{Weisz2014} comprehensively studied SFHs in the Local Group dwarf galaxies. They found that more massive systems tend to have more extended SFHs. They also showed that MW or M31 satellites have a shorter duration of star formation than those in the field populations.

Chemical abundances reflect the SFHs and nucleosynthesis pathways in satellites \citep[e.g.,][]{Tolstoy2009, Kirby2010, Kirby2011a, Kirby2011b, Ishigaki2014, Hill2019, Skuladottir2024}. \citet{Kirby2011b} analyzed metallicity distribution functions (MDFs) of the MW's satellites with Keck/DEIMOS with a chemical evolution model. They found that the MDFs of 
more-luminous systems are well-fit with their Extra Gas Model, which assumes gas infall. However, their best-fit effective yields suggested that gas outflow also played an important role in the chemical evolution of less-luminous systems.

Thanks to the difference in the delay times between core-collapse supernovae (CCSNe) and type Ia supernovae (SNe Ia), the ratios of $\alpha$-elements (e.g., Mg, Ca, and Si) to Fe are often used as an indicator for the rate of chemical evolution. For example, \citet{Hill2019} reported high-resolution spectroscopy of 99 stars in the Sculptor dSph. They found that the decreasing trend of [$\alpha$/Fe]\footnote{[X/Y] = =$\log(N_{\rm{X}}/N_{\rm{Y}})-\log(N_{\rm{X}}/N_{\rm{Y}})_{_\sun}$, where $N_{\rm{X}}$ and $N_{\rm{Y}}$ are the number densities of elements X and Y, respectively.} toward higher metallicity starts at [Fe/H] = $-1.8$. This metallicity is lower than the start of this trend in the MW, indicating that the chemical evolution of Sculptor dSph proceeded more slowly.

Numerical simulations have been performed to understand the SFHs and chemical evolution of dwarf galaxies \citep[e.g.,][]{Revaz2009, Okamoto2010, Revaz2012, Revaz2018, Hirai2015, Hirai2017a, Hirai2018, Hirai2019, Wetzel2016, Jeon2017, Escala2018, Simpson2018, Garrison-Kimmel2019, Applebaum2021, DiCintio2021, Patel2022, Samuel2022, Rodriguez2022}. \citet{DiCintio2021} found that 25\% of their simulated satellite dwarf galaxies exhibit an enhancement of star formation after infall to their host. In contrast, the star formation in satellites with little gas or small pericentric distances is quenched after infall due to ram pressure stripping. \citet{Escala2018} introduced the process of metal diffusion in cosmological zoom-in simulations of the Feedback in Realistic Environment (FIRE) project \citep{Hopkins2014}, and analyzed chemical abundances in their simulated dwarf galaxies. They found that the MDFs and intrinsic scatter in [$\alpha$/Fe] are similar in satellite and isolated dwarf galaxies, suggesting that internal chemical evolution plays a more important role than environmental effects.

Ongoing and upcoming surveys will significantly enlarge the number of stars in satellites of the MW with available spectroscopy \citep[e.g.,][]{Takada2014, Cooper2023, Jin2023}. For example, the Dark Energy Spectroscopic Instrument (DESI) Milky Way Survey will observe 7 million stars with magnitudes 16 $<r<$ 20 at Galactic latitudes $\abs{b}$ $>20^\circ$ \citep{Cooper2023}. Their footprint includes 31 Local Group dwarf galaxies. This potentially could yield medium-resolution ($R \sim 5,000$) spectroscopy of the member stars in some of these galaxies from their centers to their outskirts.

The upcoming Subaru Prime Focus Spectrograph (PFS) will target 7 Local Group dwarf galaxies in their Galactic Archaeology survey \citep{Takada2014}. Thanks to their wide field of view (1.25 square degrees) and massively multiplexed spectroscopic capability (2,394 fibers), they can obtain medium-resolution ($R \sim 5,000$) spectroscopy for stars with magnitudes $g\lesssim$ 23 in these galaxies. The Subaru PFS will yield radial velocities, [Fe/H], carbon, $\alpha$-elements, and nickel abundance measurements in each galaxy for $\approx$ 1,000 to 14,000 stars, more than ten times larger than the current numbers of stars with these measurements. Comparison with cosmological zoom-in simulations and these observations will greatly advance our understanding of the chemo-dynamical properties of dwarf galaxies.

This study aims to understand the relationship between star formation and chemical evolution in satellite galaxies. With our high-resolution cosmological zoom-in simulation of a MW-like galaxy, we examine SFHs, MDFs, and $\alpha$-element abundances in satellites with $M_{*}$ $\sim$ $10^5$--10$^7\,M_{\sun}$, corresponding to the mass ranges of satellite dSphs of the MW. We show how SFHs are reflected in MDFs and $\alpha$-element abundances using our simulation. We then evaluate the capability of upcoming surveys to reconstruct the SFHs from the chemical abundances of dwarf galaxies.

This paper is organized as follows. Section \ref{sec:methods} describes our code, the adopted initial conditions, and the procedures used for carrying out mock observations. In Section \ref{sec:results}, we describe the chemo-dynamical properties of our simulated satellites. Section \ref{sec:discussion} discusses how SFHs are reflected in chemical abundances, and how these can be observed in future surveys. Our conclusions are presented in Section \ref{sec:conclusions}.

\section{Methods} \label{sec:methods}
\subsection{Code}\label{sec:code}

We have computed the evolution of satellite galaxies in a cosmological zoom-in simulation of a MW-like galaxy performed by \citet{Hirai2022}. In this simulation, we adopted the $N$-body/density-independent smoothed particle hydrodynamics code \textsc{asura} \citep{Saitoh2008, Saitoh2009, Saitoh2013, Saitoh2016}. For cooling and heating calculations, we adopted \textsc{cloudy} ver. 13.05 \citep{Ferland2013}. Gas particles probabilistically form stars if they are in a region with a number density of hydrogen atoms higher than 100 cm$^{-3}$, the temperature is lower than 1,000 K, and there are converging flows \citep[$\nabla\cdot\bm{v}\,<\,0$, e.g.,][]{Hirai2021}.

Each star particle is treated as a simple stellar population (SSP) with the initial mass function (IMF) of \citet{Chabrier2003} from 0.1 $M_{\sun}$ to 100 $M_{\sun}$. Star particles with ages less than 10 Myr heat the surrounding gas to $10^4$ K \citep{Fujii2021}. We implemented momentum-based supernova feedback following \citet{Hopkins2018}. Metal diffusion was incorporated following \citet{Hirai2017b}. The cosmic ultra-violet (UV) heating was implemented following \citet{Haardt2012}. The reionization is assumed to {occur at} a redshift ($z$) of 8.5. We also assumed the self-shielding model of \citet{Rahmati2013}.

We adopted the nucleosynthetic yields compiled in the Chemical Evolution Library \citep[\textsc{celib},][]{Saitoh2017}. CCSNe and SNe Ia are the dominant contributors to the evolution of the [$\alpha$/Fe] ratios. For CCSNe, we used the yields of \citet{Nomoto2013} with 13 $M_{\sun}$ to 40 $M_{\sun}$. Given the mass of the star particle, we integrated the IMF from the maximum stellar mass of the IMF to the lower stellar mass until the cumulative number of stars in the integration range became unity. This approach enabled the tracking of the contribution from CCSNe with different progenitor masses in sufficiently high-resolution simulations. When the stellar particle mass ($m_{*}$) was 4.5$\,\times\,10^3\,$$M_{\sun}$, the IMF for CCSNe (13--40~$M_{\sun}$) was divided into 100 bins. For SNe Ia, we assumed a delay-time distribution with a power-law index of $-1$, and a minimum delay time of 40 Myr, following \citet{Maoz2012}. We also included the contribution of asymptotic giant branch (AGB) stars for stars with 1 to 8 $M_{\sun}$ \citep{Karakas2010, Doherty2014}. We adopted the solar abundance of \citet{Asplund2009}. {Throughout this paper, we define [$\alpha$/Fe] as the weighted average of Mg, Si, and Ca ([$\alpha$/Fe] $=$ 0.35 [Mg/Fe] $+$ 0.25 [Si/Fe] $+$ 0.40 [Ca/Fe]) to mimic [$\alpha$/Fe] measured in Keck/DEIMOS \citep{Kirby2008, Kirby2011a}.}

\subsection{Initial Conditions} \label{sec:ic}

A MW-like halo was selected from the cosmological simulation with a box size of (36 $h^{-1}$ Mpc)$^3$. We adopted cosmological parameters of $\Omega_{\rm{m}}\,=\,0.308$, $\Omega_{\Lambda}\,=\,0.692$, $\Omega_{\rm{b}}\,=\,0.0484$, and $H_{0}\,=\,67.8$ km s$^{-1}$$\>$Mpc$^{-1}$ \citep{Planck2016}. An initial condition for the zoom-in simulation was generated by \textsc{music} \citep{HahnAbel2011}. We used the Amiga Halo Finder \citep[\textsc{ahf},][]{Gill2004, Knollmann2009} to find the target halo. In this simulation, the initial masses of each particle in the finest region were $7.2\times10^4\,$$M_{\sun}$ for dark matter, $1.3\times10^4\,$$M_{\sun}$ for gas, and $4.5\times10^3\,$$M_{\sun}$ for stars. We set the gravitational softening length ($\epsilon_{\rm{g}}$) to 85 pc for dark matter and 82 pc for gas and stars. We performed the simulation from $z$ = 100 to 0.

In this simulation, we picked out satellites orbiting the central galaxy. We only considered those with a minimum of 10$^4$ dark matter and 10 star particles, and made sure that they were not false substructures introduced by the contamination from low-resolution particles. Table \ref{tab:list} lists the simulated satellite galaxies selected for this study.

\begin{deluxetable}{cccccr}
\tablecaption{List of Simulated Satellite Galaxies at $z$ = 0.\label{tab:list}}
\tablewidth{0pt}
\tablehead{
    \colhead{Halo ID} & \colhead{$M_{\rm{halo}}$} & \colhead{$M_*$} & \colhead{$\langle {\rm [Fe/H]} \rangle$} & \colhead{$\sigma_{\rm{[Fe/H]}}$ }& \colhead{$d$}\\
    \colhead{}&\colhead{($M_{\sun}$)}&\colhead{($M_{\sun}$)}&\colhead{}&\colhead{(dex)}&\colhead{(kpc)}}
\decimals
\startdata		
            9 & $7.5\times10^9$ & $7.5\times10^6$ & $-1.95$ & 0.23 & 204.2\\
		12 & $4.7\times10^9$ & $2.1\times10^7$  & $-1.08$ & 0.58 & 148.8\\
		36 & $2.2\times10^9$ & $1.1\times10^7$  & $-1.43$ & 0.37 & 54.5\\
  		38 & $2.5\times10^9$ & $1.3\times10^5$  & $-1.52$ & 0.52 & 198.7\\
		40 & $2.3\times10^9$ & $3.9\times10^6$  & $-1.52$ & 0.46 & 57.9\\
  		150 & $5.9\times10^8$ & $2.4\times10^5$  & $-2.53$ & 0.24 & 190.7\\
            151 & $6.0\times10^8$ & $7.7\times10^4$  & $-2.89$ & 0.43 & 167.2\\
		167 & $5.2\times10^8$ & $3.1\times10^4$  & $-4.34$ & 0.14 & 206.6\\
		199 & $4.2\times10^8$ & $2.8\times10^4$  & $-3.42$ & 0.28 & 169.2\\
\enddata
\tablecomments{From left to right, the columns are the Halo ID, the total halo mass within the virial radius ($M_{\rm{halo}}$), the total stellar mass ($M_*$), the mean [Fe/H] ($\langle {\rm [Fe/H]} \rangle$), the dispersion of [Fe/H] ($\sigma_{\rm{[Fe/H]}}$), and the distance from the center of the central galaxy ($d$). $M_*$, $\langle {\rm [Fe/H]} \rangle$, and $\sigma_{\rm{[Fe/H]}}$ are computed within the half-mass radius.}
\end{deluxetable}

\subsection{Mock Observations} \label{sec:mock}

We performed mock observations for Subaru PFS (see Section \ref{sec:future})\footnote{\citet{Sanderson2020} also discussed in detail mock observations of galaxy simulations.}. For the mock observation, we computed the magnitudes of simulated stars. First, SSP particles were divided into individual stars. In this model, stars from 0.1 $M_{\sun}$ to 100 $M_{\sun}$ were probabilistically generated from SSP particles, following a \citet{Chabrier2003} IMF. Stars were generated until the total generated stellar mass exceeded the particle's mass. Then, the magnitudes of each star were computed using the isochrone table obtained from \textsc{cmd} 3.7\footnote{\url{http://stev.oapd.inaf.it/cgi-bin/cmd}} \citep[][and updates thereof]{Girardi00}. 

We generated isochrones with ages from 4 Gyr to 13.8 Gyr and [M/H]\footnote{[M/H] = $\log(Z/X)-\log(Z/X)_{\sun}$, where $X$ and $Z$ are the mass fractions of hydrogen and metals, respectively.} from $-2.0$ to 0.0 based on the PARSEC-COLIBRI stellar-evolutionary tracks \citep{Bressan2012, Chen2014, Chen2015, Tang2014,Marigo2017,Pastorelli2019,Pastorelli2020}. With this tool, we computed apparent $V$-band magnitudes for stars in Halos 12 and 40. We assume Halos 12 and 40 are located at 147 kpc and 86 kpc from an observer to compare with the Fornax and Sculptor dSphs, respectively \citep{Mcconnachie2012}.

We then applied the Subaru PFS spectral synthesis pipeline \citep[roughly based on][]{Kirby2010, Escala2019} to compute observed uncertainties. The pipeline adopts synthetic spectra of K-giants and G-dwarfs for $-4.0 \leq$ [Fe/H] $\leq -0.5$. We calculated wavelength-dependent continuum signal-to-noise ratios with the Subaru PFS Exposure Time Calculator\footnote{\url{https://github.com/Subaru-PFS/spt_ExposureTimeCalculator}} using the simulated stars’ $V$-band magnitudes, assuming a three-hour exposure in the Subaru PFS median-resolution mode for K giants. Then, we computed uncertainties on [Fe/H] and [$\alpha$/Fe] by resampling the synthetic spectra hundreds of times from Gaussian-distributed per-pixel noise inversely proportional to the estimated signal-to-noise ratios.  The simulated chemical abundances of stars are varied within those estimated uncertainties.

\section{Results} \label{sec:results}

\subsection{Structures and Star Formation Histories} 
\label{sec:sfh}

This paper mainly discusses the chemo-dynamical evolution of Halos 12, 40, and 150, listed in Table \ref{tab:list}. The [$\alpha$/Fe] as a function of [Fe/H] for Halos 9 and 36 are shown in the Appendix. We select three these simulated dwarf galaxies based on their stellar mass (Halo 12: $2.1\times10^7M_{\sun}$, Halo 40: $3.9\times10^6M_{\sun}$, and Halo 150: $2.4\times10^5M_{\sun}$). These values are similar to those of the Fornax ($2.0\times10^7 M_{\sun}$), Sculptor ($2.3\times10^6 M_{\sun}$), and Draco ($2.9\times10^5 M_{\sun}$) dSphs \citep{Mcconnachie2012}. Also, Halos 12, 40, and 150 currently contain no gas.

\begin{figure}[ht!]
\epsscale{1.1}
\plotone{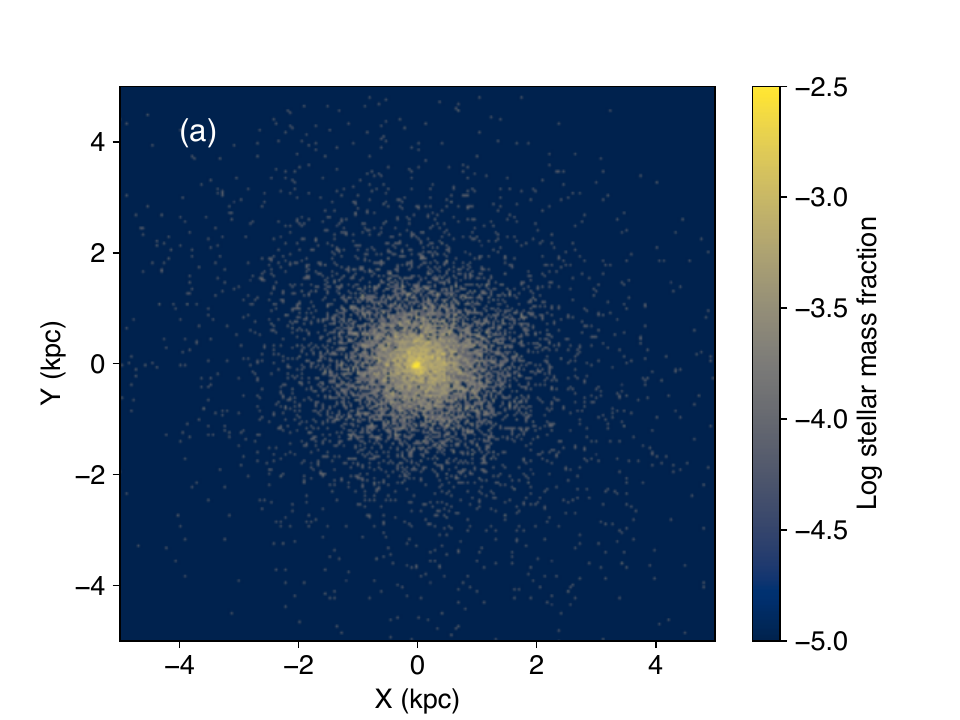}
\plotone{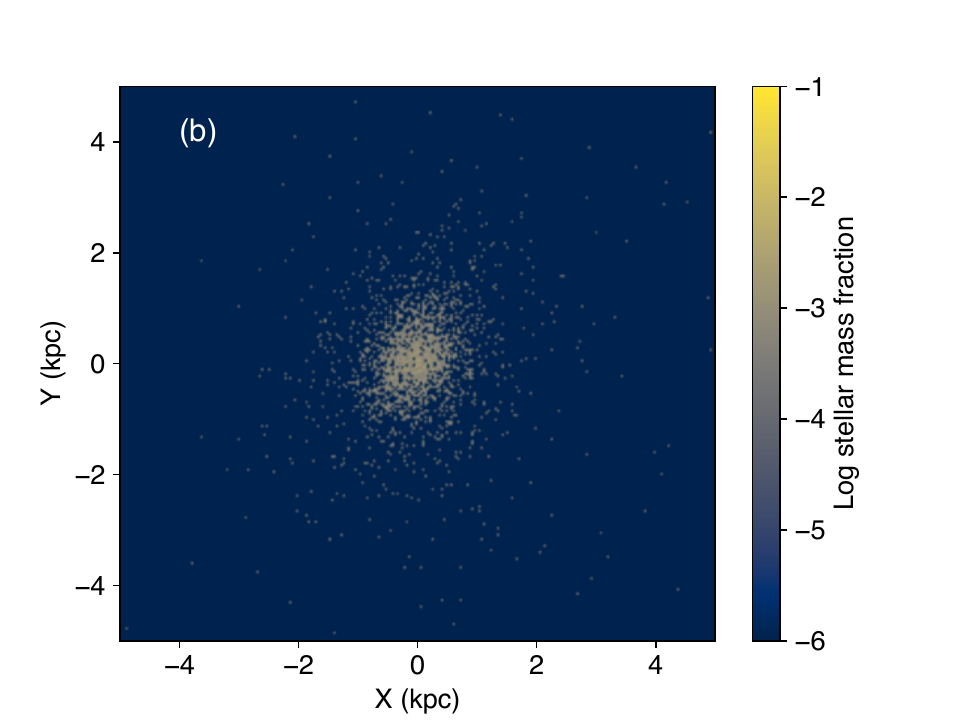}
\plotone{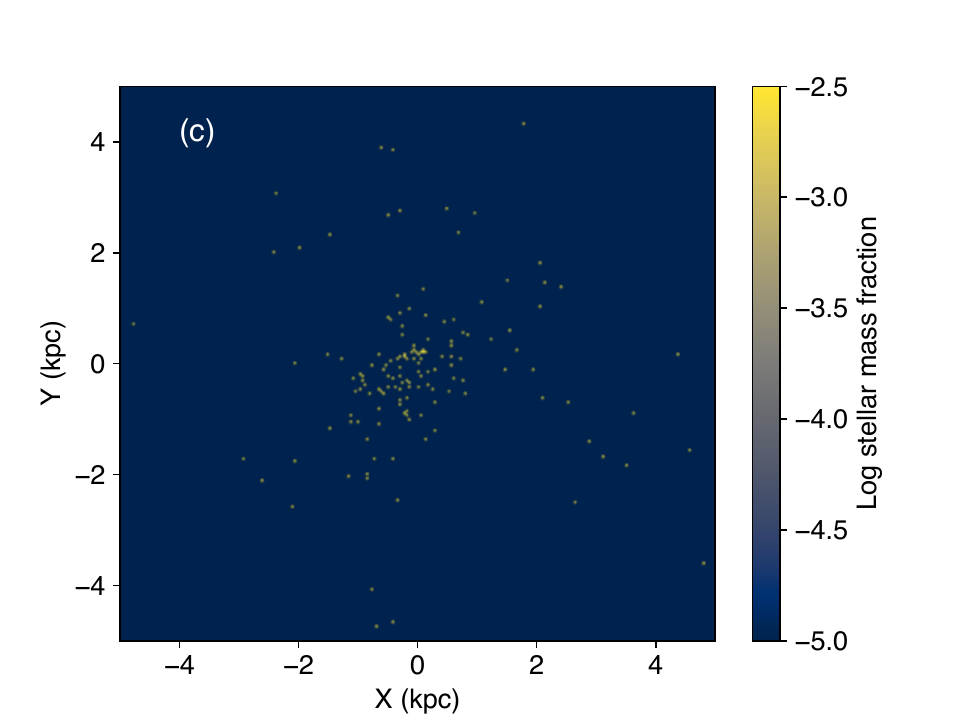}
\caption{Stellar distribution of simulated satellite dwarf galaxies  for (a) Halo 12, (b) Halo 40, and (c) Halo 150. The color scale depicts each grid's log scale stellar-mass fraction. {The stellar mass fraction is defined as $M_{*,\rm{grid}}/M_{*,\rm{tot}}$, where $M_{*,\rm{grid}}$ and $M_{*,\rm{tot}}$ are the stellar mass contained in each grid and total stellar mass of the galaxy, respectively. The grid size is 0.08 $\times$ 0.08 kpc$^2$.} Most stars are spherically distributed at the center of their dark matter halo.
\label{fig:stars}}
\end{figure}

Figure \ref{fig:stars} shows the stellar mass distribution of Halos 12, 40, and 150 at $z$ = 0. The half-mass (light) radii of these galaxies are 1,334 pc (Halo 12), 874 pc (Halo 40), and 1,346 pc (Halo 150), respectively. The somewhat larger radii than the observed ones \citep[Fornax: 710 pc, Sculptor: 283 pc, Draco: 221 pc,][]{Mcconnachie2012} are due to the spatial resolution of this simulation ($\epsilon_{\rm{g}}=85$ pc).


\begin{figure}[ht!]
\epsscale{1.3}
\plotone{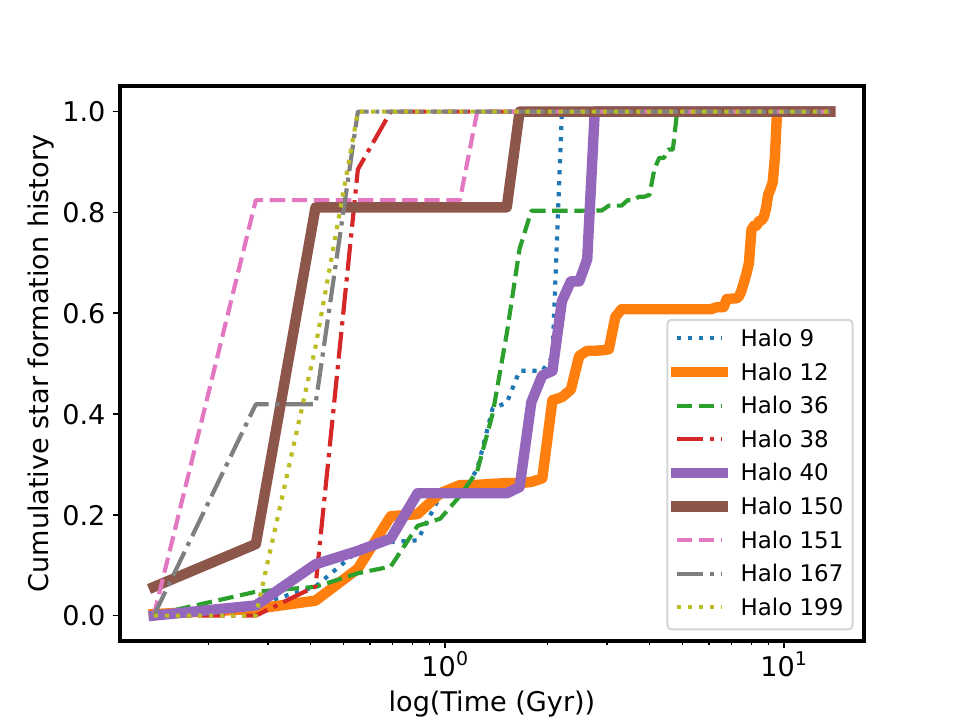}
\caption{Cumulative SFHs of simulated dwarf satellites, as listed in Table \ref{tab:list}. Less massive halos (e.g., Halos 151, 167, and 199) tend to quench star formation earlier than more massive halos (e.g., Halos 9, 12, and 36). {The cumulative SFH is defined as $M_{*}(t)/M_{*}(T)$, where $M_{*}(t)$ and $M_{*}(T)$ are the stellar masses at the cosmic time $t$ and the present ($z$ = 0), respectively.}\label{fig:csfh}}
\end{figure}

The simulated satellite dwarf galaxies exhibit various SFHs. Figure \ref{fig:csfh} shows the cumulative SFHs of all satellites listed in Table \ref{tab:list}. The SFHs of satellite galaxies are affected by SN feedback, cosmic reionization, and interactions with the host galaxy. This figure shows that more massive satellites tend to have extended SFHs, while less massive halos quench star formation earlier. Star formation in halos with $<10^9M_{\sun}$ (150, 151, 167, and 199) is quenched at $<$ 2 Gyr from the beginning of the simulation by cosmic reionization and SN feedback, while halos with $\geq\,10^9M_{\sun}$ form stars after the reionization epoch. Gas accreted before reionization in halos with $\geq\,10^9M_{\sun}$ self-shield the UV background, resulting in them surviving the reionization \citep[e.g.,][]{Onorbe2015, Wheeler2019}. Hereafter, we focus on three satellites: Halos 12, 40, and 150.

\begin{figure}[ht!]
\epsscale{1.2}
\plotone{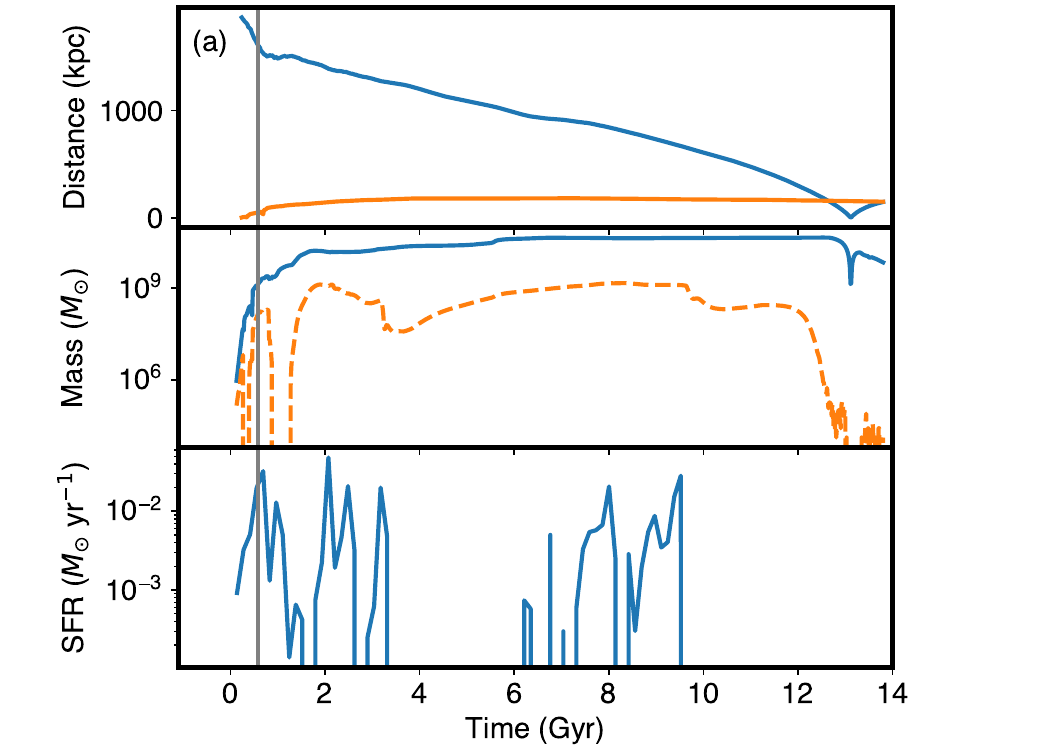}
\plotone{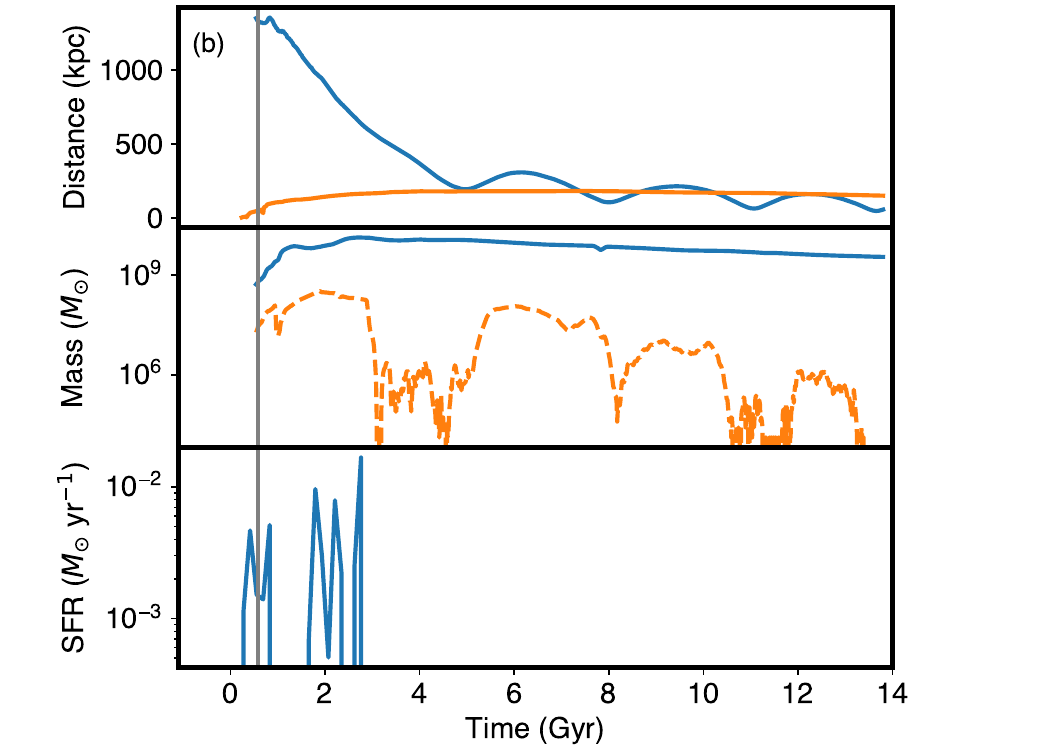}
\plotone{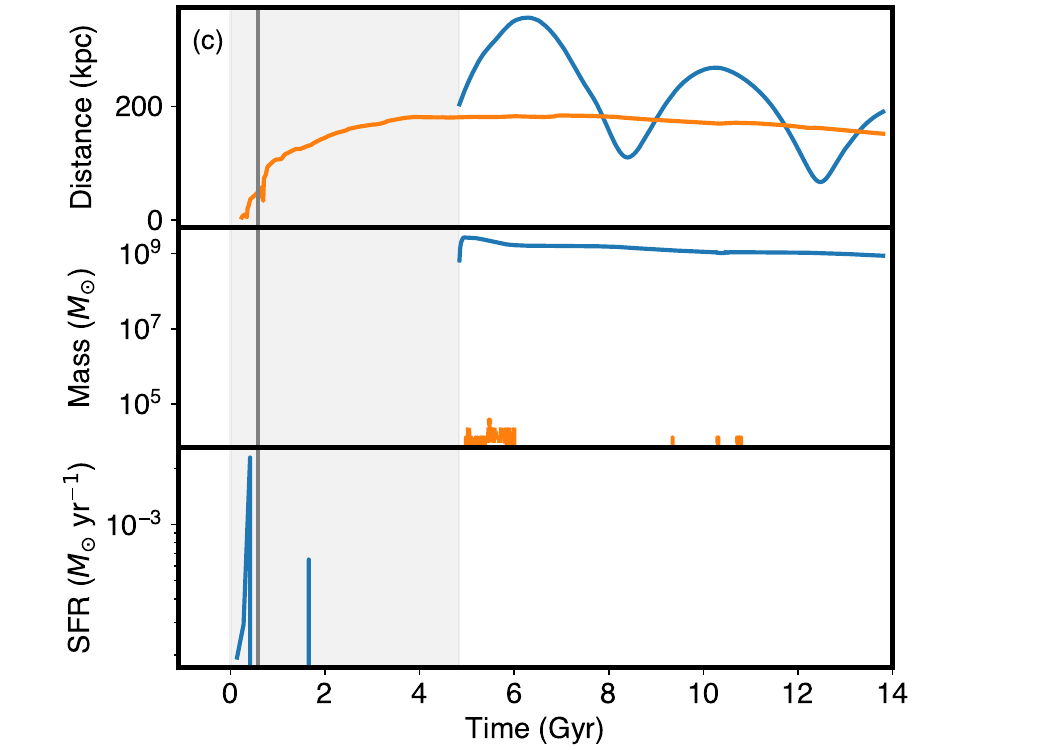}
\caption{Cosmic time evolution of (a) Halo 12, (b) Halo 40, and (c) Halo 150. Top sub-panels: The orbital distance (blue) and the time evolution of the virial radius of the main halo (orange). Middle sub-panels: The dark matter (blue-solid) and gas (orange-dashed) mass evolution. Bottom sub-panels: SFHs. The grey line represents the epoch of reionization ($z = 8.5$). The light-grey shaded region in panel (c) means the halo finder cannot follow the mass evolution.\label{fig:sfh}}
\end{figure}

The mass and the cosmic infall time also affect the SFHs. Figure \ref{fig:sfh} shows the orbits (top panels), mass evolution (middle panels), and SFHs (bottom panels) of Halos (a) 12, (b) 40, and (c) 150. Halo 12 has the most recent infall time. The first pericentric passage (5 kpc) of this galaxy is 0.7 Gyr prior to the end of the simulation (Figure \ref{fig:sfh} (a), top panel). Prior to pericentric passage, this galaxy experienced two star formation events separated by 2.9 Gyr (Figure \ref{fig:sfh} (a), bottom panel). The first period of star formation starts at 0.1 Gyr and ends at 3.3 Gyr from the beginning of the simulation. During this period, stars are formed along with the accretion of material (Figure \ref{fig:sfh} (a), middle panel). After SNe expel the gas away from the halo, the infall of the gas forms new stars. This interplay episodically forms stars for 3.2 Gyr. 

The second star formation event begins when the accretion of a halo brings additional material to the halo at 6.2 Gyr. As with the first period of star formation, it is regulated by SN feedback. The star formation is quenched when feedback from CCSNe from the recent star formation ($t\lesssim$ 10 Myr ago) and SNe Ia from previous star formation ($t$ $\sim$ 1 Gyr ago) expel the gas from the galaxy at 9.5 Gyr.

Halo 40 has a shorter total duration of star formation, mainly due to the earlier infall time than that of Halo 12. Halo 40 crosses the main halo's virial radius ($R_{\rm{vir}}$) at 7.4 Gyr, while Halo 12 experiences its closest pericenter passage at 12.6 Gyr. Due to the early infall, repeated gas removal by ram pressure stripping prevents additional star formation in the later phase. The evolution of gas mass after the first infall is due to our analysis method. The increase in the tidal radius of the halo around the apocenter accretes more diffuse gas around the galaxy, resulting in the increase of the detected gas mass of this halo. Although gas mass evolution is shown here, these gas particles are not eligible to form stars.

Halo 40 experienced star formation in the first 2.8 Gyr. As shown in the bottom panel of Figure \ref{fig:sfh} (b), there are five peaks of star formation, separated from 0.40 to 0.97 Gyr. The SFH in this halo is also mainly regulated by SN feedback. As shown in the bottom panel of Figure \ref{fig:sfh} (b), stars are formed during cosmic reionization. After the star formation is quenched at 0.83 Gyr, an additional gas supply resumes star formation at 1.79 Gyr. Eventually, star formation is halted at 2.76 Gyr. This quenching is mainly caused by the heating by CCSNe from the recent star formation and SNe Ia from the previous star formation, due to their delay times. Since Halo 40 is located at a distance five times larger than the virial radius of the main halo at 2.76 Gyr, ram pressure stripping is unlikely to be the main cause responsible for the suppression of star formation.

Halo 150 has the shortest duration of star formation among the halos shown in Figure \ref{fig:sfh}.  Figure \ref{fig:sfh} (c) shows the cosmic time evolution of Halo 150. The top panel shows that this halo experienced at least two pericenter passages. Note that we cannot follow the mass evolution before 4.84 Gyr, because the progenitor halos are undetected by the halo finder. As shown in the bottom panel of Figure \ref{fig:sfh} (c), the first episode of star formation lasts 0.47 Gyr, and is quenched by cosmic reionization. In this episode, 80\% of its stars are formed. The second star formation event occurs at 1.66 Gyr, possibly because of the gas infall, but it is quenched quickly.

\subsection{Chemical Abundances}\label{sec:chem}

\begin{figure}[ht!]
\epsscale{1.2}
\plotone{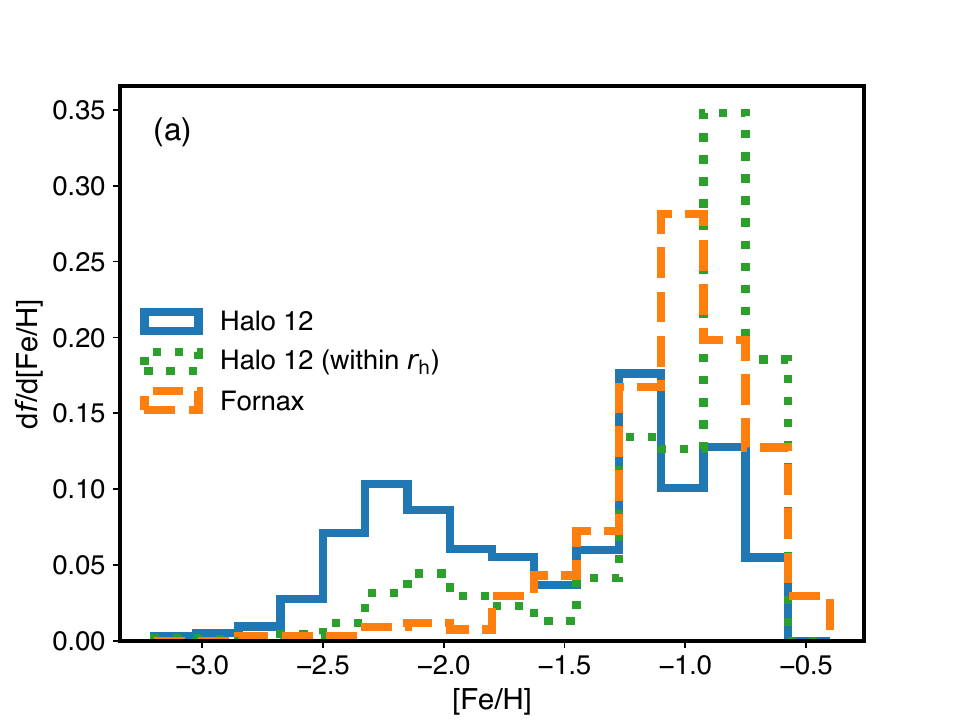}
\plotone{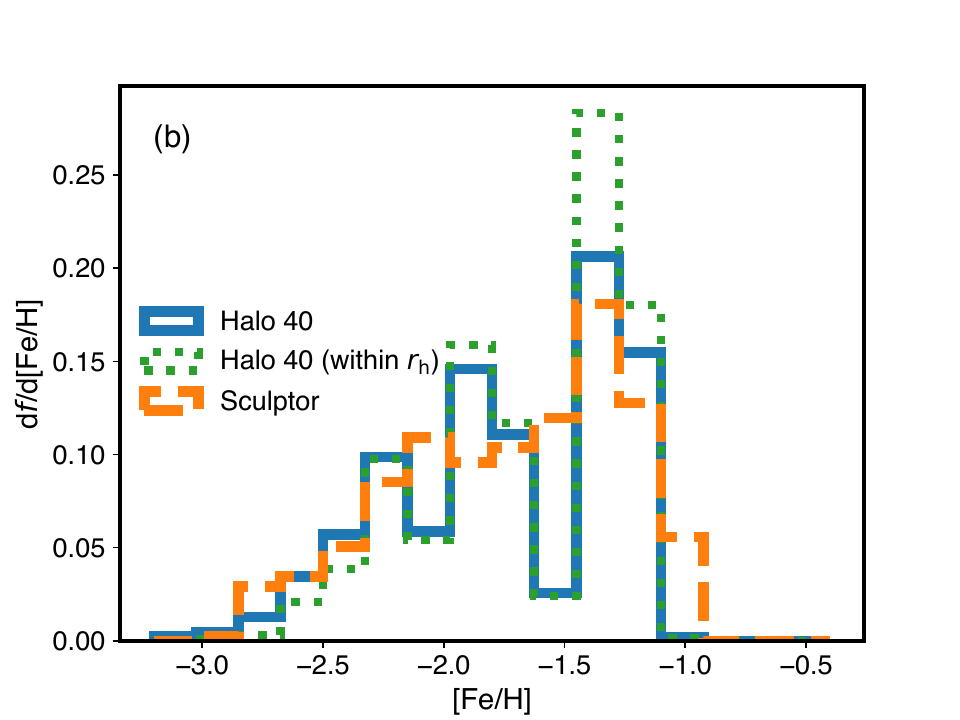}
\plotone{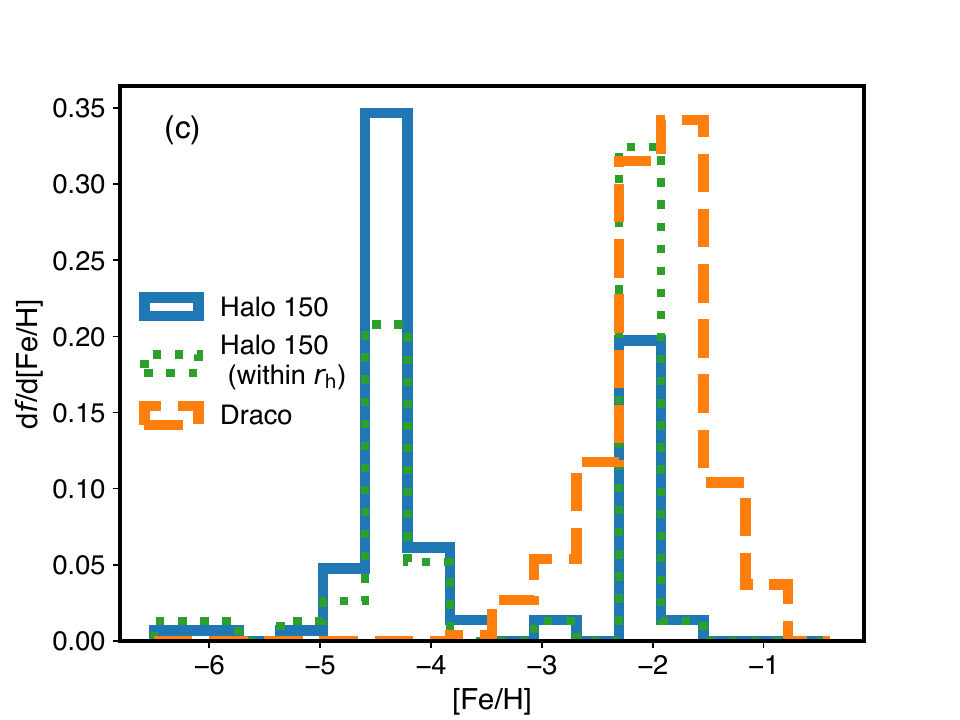}
\caption{Simulated (blue-solid line) and observed (orange-dashed line) MDFs for (a) Halo 12 and Fornax, (b) Halo 40 and Sculptor, and (c) Halo 150 and Draco. The green-dashed line represents the MDFs for stars within $r_{\rm{h}}$. The simulated data do not include simulated observational errors.  Observed data are taken from \citet{Kirby2010}. \label{fig:mdf}}
\end{figure}

The MDFs of stellar systems reflect their histories of star formation, gas infall, and gas outflow; Figure \ref{fig:mdf} shows MDFs of Halos 12, 40, and 150. We also plot the observed MDFs of the Fornax, Sculptor, and Draco dSphs \citep{Kirby2010}. It should be noted that the purpose of our study is {\it not} to reproduce the MDFs of the observed dSphs. Rather, we compare simulated and observed MDFs in Section \ref{sec:chemo-dynamics}.

The MDF of Halo 12 exhibits a bimodal distribution, reflecting two major star formation events (the bottom panel of Figure \ref{fig:sfh} (a)). All stars with [Fe/H] $< -1.5$ are formed within 3.3 Gyr from the beginning of the simulation. These stars are mainly located in the outskirts of the galaxy. For stars with [Fe/H] $< -1.5$, 28.5\% of them are within $r_{\rm{h}}$, while 71.5\% of stars with [Fe/H] $\geq -1.5$ are within $r_{\rm{h}}$. As shown in the green-dashed line in Figure \ref{fig:mdf} (a), the fraction of stars with [Fe/H] $< -1.5$ in the MDF is significantly decreased for stars within $r_{\rm{h}}$. Stars around [Fe/H] = $-1.2$ and [Fe/H] = $-0.8$ are associated with star formation events around 8.0 Gyr and 9.5 Gyr, respectively. As shown in the middle panel of Figure \ref{fig:sfh} (a), these stars are formed from gas infall.

Figure \ref{fig:mdf} (b) shows the MDF of Halo 40. The MDF is broadly distributed over 
$-3.0\lesssim$ [Fe/H] $\lesssim-1.0$. Stars around [Fe/H] = $-$2.3, $-$1.8, and $-$1.3 reflect star formation at different cosmic times. For [Fe/H] $< -$2.5, all stars formed before 1.0 Gyr from the beginning of the simulation. For stars with $-$2.5 $<$ [Fe/H] $<$ $-$2.0, half of them are formed at $t\,<$ 1.0 Gyr, while others are formed at 1.7 $<$ $t$/(Gyr) $<$ 2.3, simultaneously with stars with $-$2.0 $<$ [Fe/H] $<$ $-$1.5. Stars with [Fe/H] $>-$1.5 have younger ages. All of these stars are formed after 2.2 Gyr. Although there is an overlap in the ages of each peak, the peaks in the MDFs indicate star formation at different cosmic times. In Figure \ref{fig:mdf} (b), we also plot the MDF for stars within $r_{\rm{h}}$. Unlike for Halo 12, the MDFs are not largely affected by the spatial distribution of stars.

Figure \ref{fig:mdf} (c) shows the MDF for Halo 150. As shown in Figure \ref{fig:sfh} (c), Halo 150 exhibits two 
star formation events. The second peak of star formation produces stars with $-$2.3 $<$ [Fe/H] $<$ $-$2.1, while the first star formation event mainly forms stars with [Fe/H] $< -$4.0. The number fraction of these ultra metal-poor (UMP) stars is 75.6\% for all UMP stars and 64.9\% for UMP stars within $r_{\rm{h}}$. These stars largely affect the median metallicity of this galaxy. The median metallicity is [Fe/H] = $-$4.36 for all stars, but [Fe/H] = $-$2.16 for stars with [Fe/H] $> -$4.0.

\begin{figure}[ht!]
\epsscale{1.2}
\vspace{-1.5cm}
\plotone{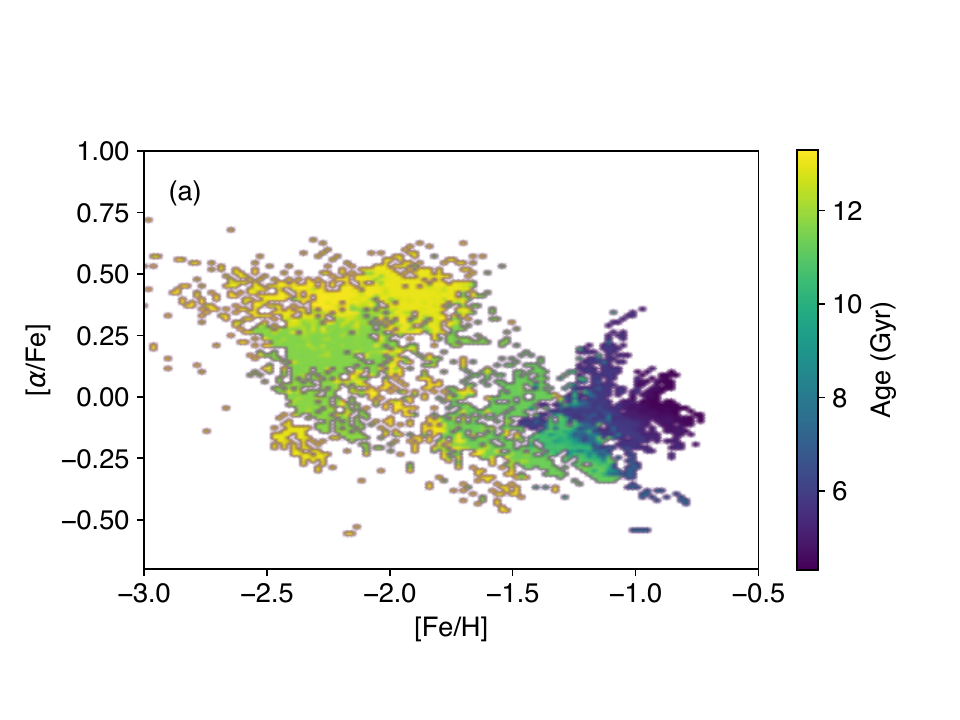}
\vspace{-1.2cm}
\vspace{-1.2cm}
\plotone{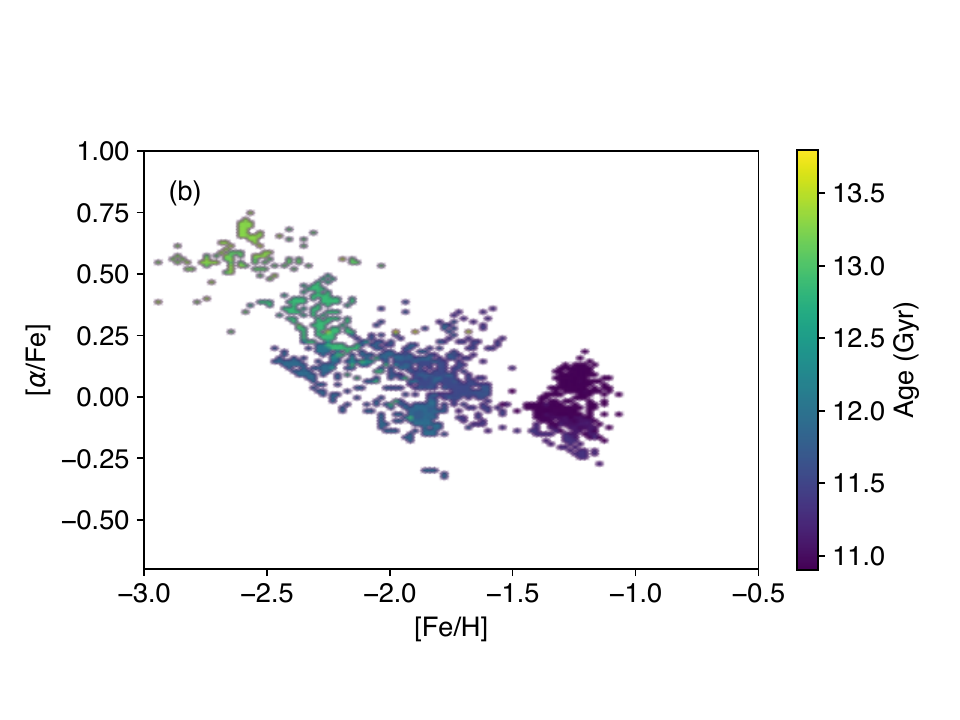}
\vspace{-1.2cm}
\vspace{-1.2cm}
\plotone{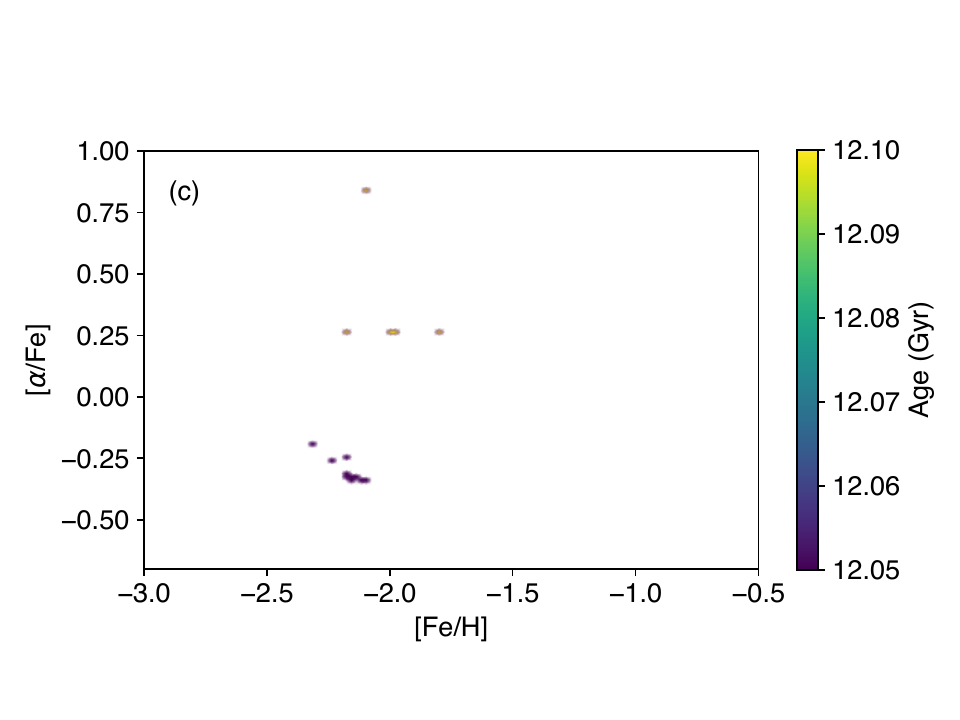}
\vspace{-1.2cm}
\caption{The $\alpha$-element distributions for (a) Halo 12, (b) Halo 40 , and (c) Halo 150.  
The color bars indicate the ages of the stars.  The simulated data do not include simulated observational errors.\label{fig:alpha}}
\end{figure}

Stars with different ages clearly differ in the [$\alpha$/Fe] vs. [Fe/H] space. Figure \ref{fig:alpha} (a) shows [$\alpha$/Fe], as a function of [Fe/H], in Halo 12. This galaxy has two major star formation events (Figure \ref{fig:sfh} (a)). The first event (13.7 Gyr to 10.5 Gyr ago) forms the decreasing trend of [$\alpha$/Fe] from [Fe/H] = $-2.5$ to [Fe/H] = $-1.0$. Also, there is roughly a $\sim$1\,dex scatter in the [$\alpha$/Fe] ratios. The episodic star formation creates these features during the first major star formation event. The first star formation episode ($\geq$ 13 Gyr ago) forms the high-$\alpha$ ([$\alpha$/Fe] $>+$0.3) component. The interstellar medium (ISM)'s inhomogeneity results in a widely distributed metallicity ($-3.0<$ [Fe/H] $<-1.5$). The low-$\alpha$ ($-0.3<$ [$\alpha$/Fe] $<-0.1$) and very metal-poor ($-2.5<$ [Fe/H] $<-2.2$) component come from another dwarf galaxy accreted to Halo 12. The subsequent star formation episodes (12.0 Gyr to 10.5 Gyr ago) produce the decreasing trend of [$\alpha$/Fe] ratios due to the substantial contribution from SNe Ia.

In contrast, the second star formation event (7.6 Gyr to 4.3 Gyr ago) produces an increasing trend of the [$\alpha$/Fe] ratios for [Fe/H] $>-1.5$. This trend suggests that stars are preferentially formed from the ejecta of CCSNe. During the second major star formation event, stars are mainly produced at the galaxy's center. Young stars give rise to CCSNe mainly at the center, while SNe Ia occur in the more extended region due to their delay times; SNe Ia occur in the more distant places relative to the star-forming region. This difference in the spatial distribution results in the formation of stars reflecting the yields of CCSNe. Since Si also exhibits a similar behavior, AGB stars are unlikely to contribute to forming this trend.

Figure \ref{fig:alpha} (b) shows [$\alpha$/Fe], as a function of [Fe/H], in Halo 40. From inspection, five peaks of star formation (Figure \ref{fig:sfh} (b)) produce groups of stars with different [Fe/H] and [$\alpha$/Fe] ratios. The first peak of star formation (13.4 Gyr ago) produces stars with [Fe/H] $< -$2.3 and [$\alpha$/Fe] $>+$0.3. Since it is the earliest phase of the star formation, CCSNe are the dominant contributor to the enrichment, resulting in a flat trend of [$\alpha$/Fe] as a function of [Fe/H]\@. A few stars with [Fe/H] $>-$2.0 and [$\alpha$/Fe] $\approx$ 0.2 are formed from the ejecta of Population III CCSNe. The second peak of star formation (13.0 Gyr ago) forms stars with $-2.5<$ [Fe/H] $< -$2.0 and $+0.1<$ [$\alpha$/Fe] $<+$0.5. The contribution of SNe Ia from the stars produced in the first peak of star formation makes this second group of stars, with lower [$\alpha$/Fe] and higher [Fe/H] than the first group.

Subsequent star formation and the contributions of SNe Ia from the previous peaks of star formation produce groups of stars with lower [$\alpha$/Fe] and higher [Fe/H]\@. The third peak of star formation (12.0 Gyr ago) creates groups of stars with $-2.5<$ [Fe/H] $< -1.7$ and $-0.3<$ [$\alpha$/Fe] $<+0.2$. This group has the lowest [$\alpha$/Fe] ratios because of the contribution of SNe Ia from the previous two star formation peaks. The fourth peak of star formation (11.6 Gyr ago) produces stars with the same [Fe/H] range but higher [$\alpha$/Fe] ratios (0.0 $<$ [$\alpha$/Fe] $<+0.4$). This group of stars reflects the ejecta from CCSNe formed in the third peak of star formation. The final star formation event (11.0 Gyr ago) forms stars with $-1.5<$ [Fe/H] $< -1.0$ and $-0.2<$ [$\alpha$/Fe] $<+0.2$. Because of its short duration ($\sim$100 Myr), stars are mainly formed from the ejecta of CCSNe.

Figure \ref{fig:alpha} (c) shows [$\alpha$/Fe] as a function of [Fe/H] in Halo 150. Although stars are too few to discuss the trend of the [$\alpha$/Fe] ratios, stars formed at different times exhibit distinct differences in [$\alpha$/Fe] ratios. Stars formed in $\geq$ 13.4 Gyr ago show [$\alpha$/Fe] $>+0.2$, reflecting the yields of CCSNe. Different [$\alpha$/Fe] ratios originate from CCSNe with different progenitor masses. A clear separation of star formation events (1.24 Gyr, Figure \ref{fig:sfh} (c)) yields stars formed in the second star formation peak with lower [$\alpha$/Fe] ratios owing to the contribution of SNe Ia.

The dispersion of the [$\alpha$/Fe] ratios reflects the degree of the ISM's inhomogeneity. We quantified the scatter for [$\alpha$/Fe] in $-3<$ [Fe/H] $< -0.5$  following \citet{Escala2018}. These authors defined the intrinsic scatter as the standard deviation of the distance distribution between stars' [$\alpha$/Fe] ratios and the cubic spline fitting curve for the data. For Halos 12 and 40, the intrinsic scatter of [$\alpha$/Fe] is 0.18\,dex  and 0.16\,dex, respectively. These are similar to the estimated intrinsic scatter \citep{Escala2018} of the Fornax (0.14\,dex) and Sculptor dSphs (0.078 \,dex), meaning that the simulated and observed satellites have ISM inhomogeneity that gives rise to scatter $\lesssim 0.2$ dex\,for the [$\alpha$/Fe] ratios.

\begin{figure*}[ht!]
\epsscale{1.17}
\plottwo{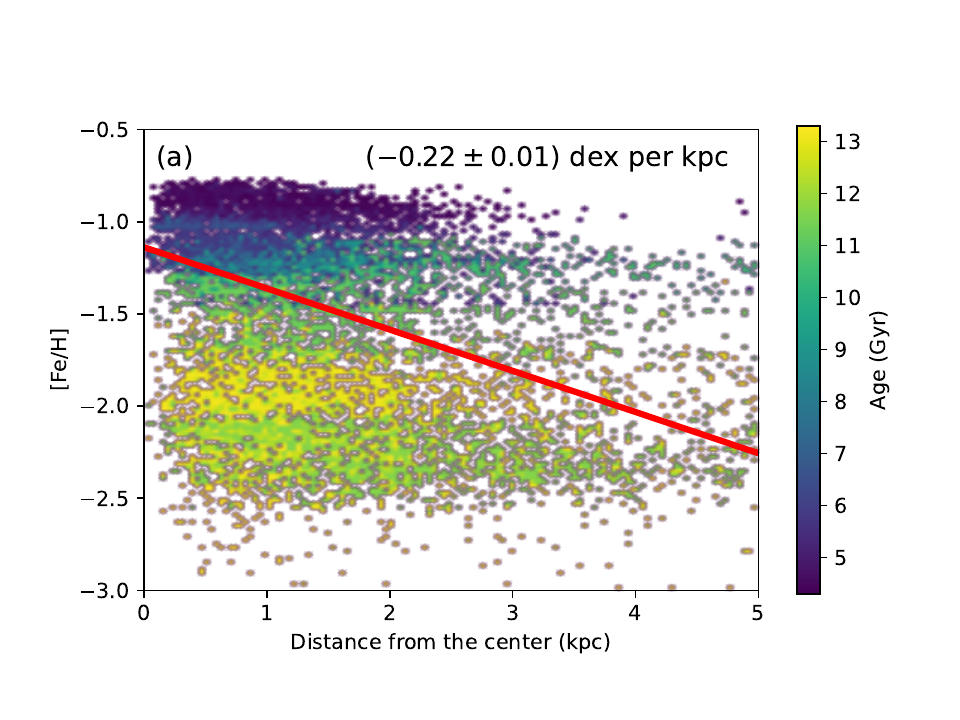}{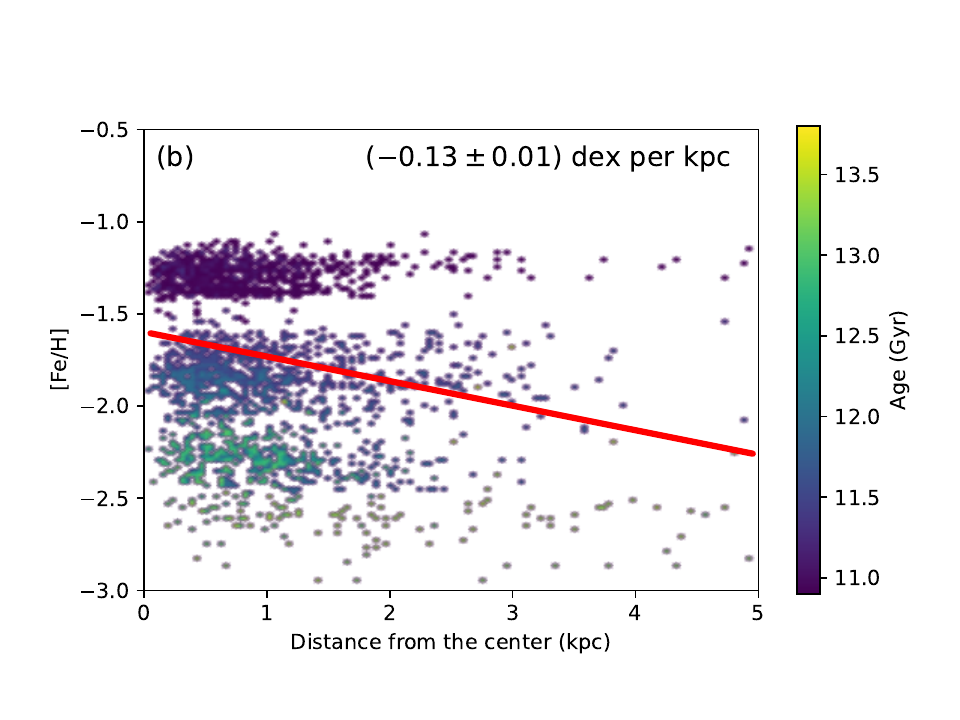}
\plottwo{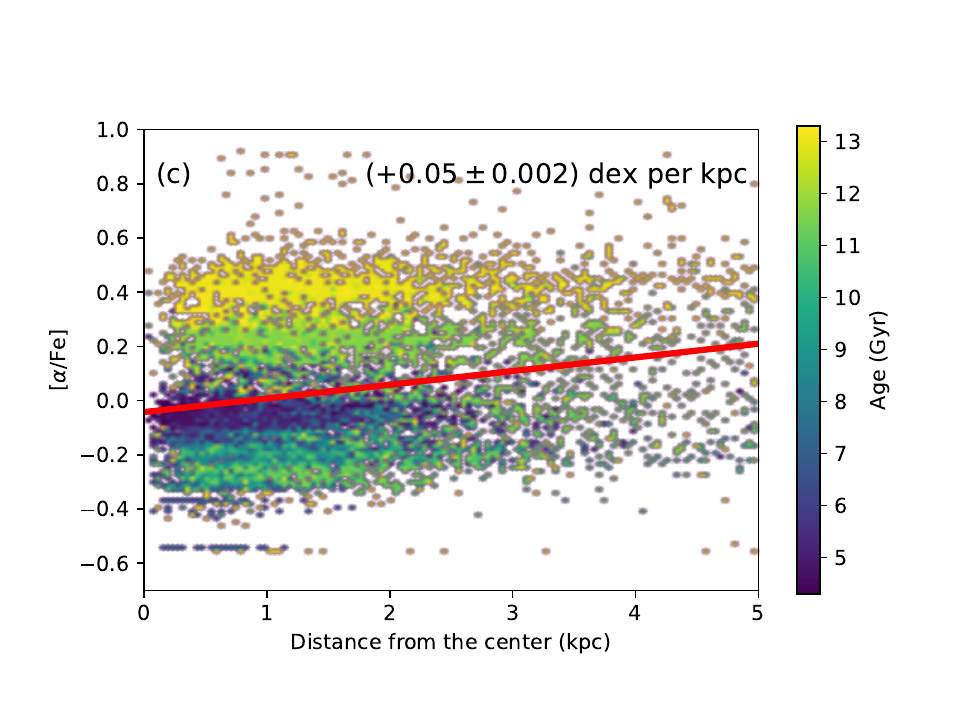}{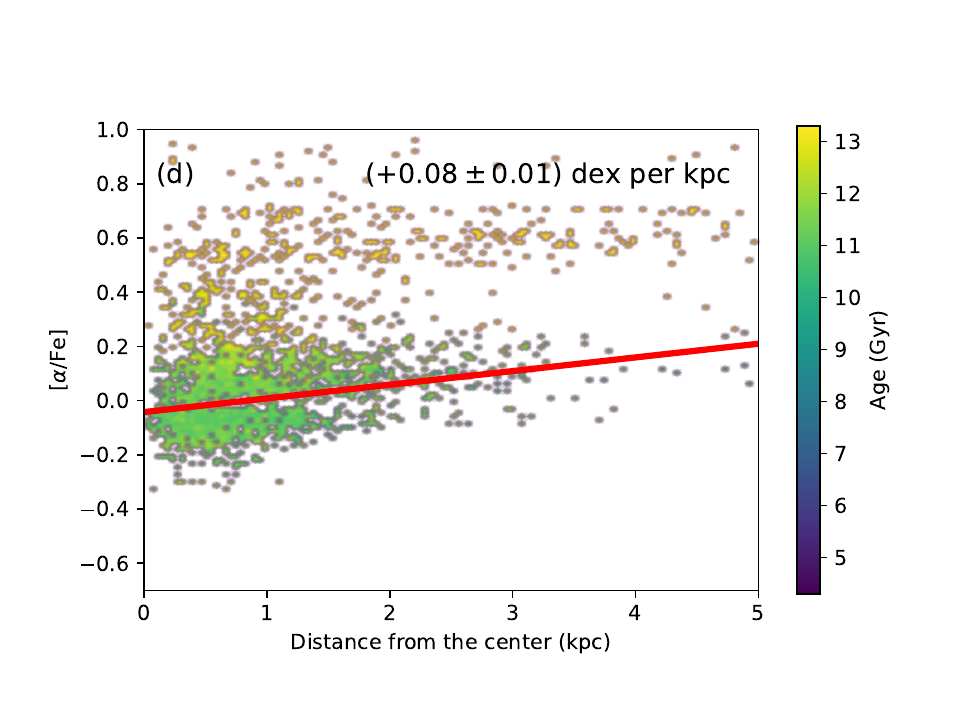}
\caption{Radial [Fe/H] distributions for (a) Halo 12, (b) Halo 40, and [$\alpha$/Fe] distributions for (c) Halo 12, and (d) Halo 40, respectively. The color bars indicate the ages of the stars. The simulated data do not include simulated observational errors. The red line is the least squares linear fit for the data. The slope is shown in each panel.\label{fig:radial}}
\end{figure*}

The radial metallicity distribution reflects spatial variations in star formation. Star formation in the inner region of Halos 12 and 40 lasts longer than that in the outer region. Figures \ref{fig:radial} (a) and (b) show radial [Fe/H] distributions in Halos 12 and 40, respectively. Both galaxies have a negative slope of [Fe/H], as a function of the distance from the center, reflecting the difference in the spatial distribution of the stars with different ages. The youngest stars in these galaxies are located within 3 kpc, while stars with ages of $>$ 13 Gyr have a more extended spatial distribution to 5 kpc.

The radial [$\alpha$/Fe] distribution exhibits positive slopes (Figures \ref{fig:radial} (c) and (d)). Because newer stars located in the center of the galaxies are more affected by SNe Ia, the average [$\alpha$/Fe] ratio near the galactic center is lower than in the outskirts. These radial [Fe/H] and [$\alpha$/Fe] gradients are caused by old and metal-poor populations in the outskirts. This result highlights the importance of measuring the chemical abundances of stars in the outer regions of dwarf satellites.

\begin{figure}[ht!]
\epsscale{1.1}
\plotone{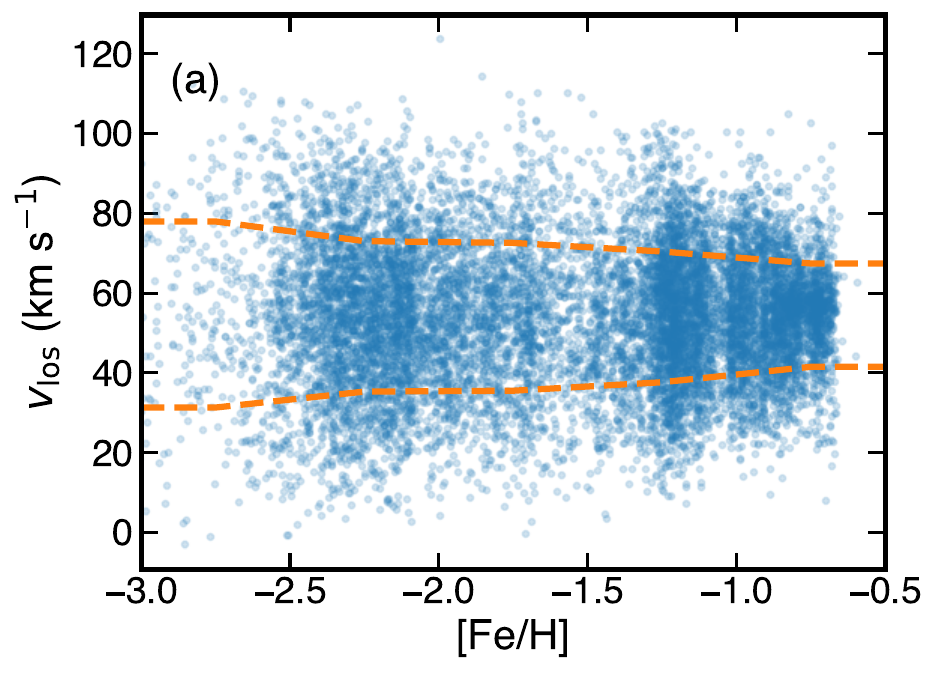}
\plotone{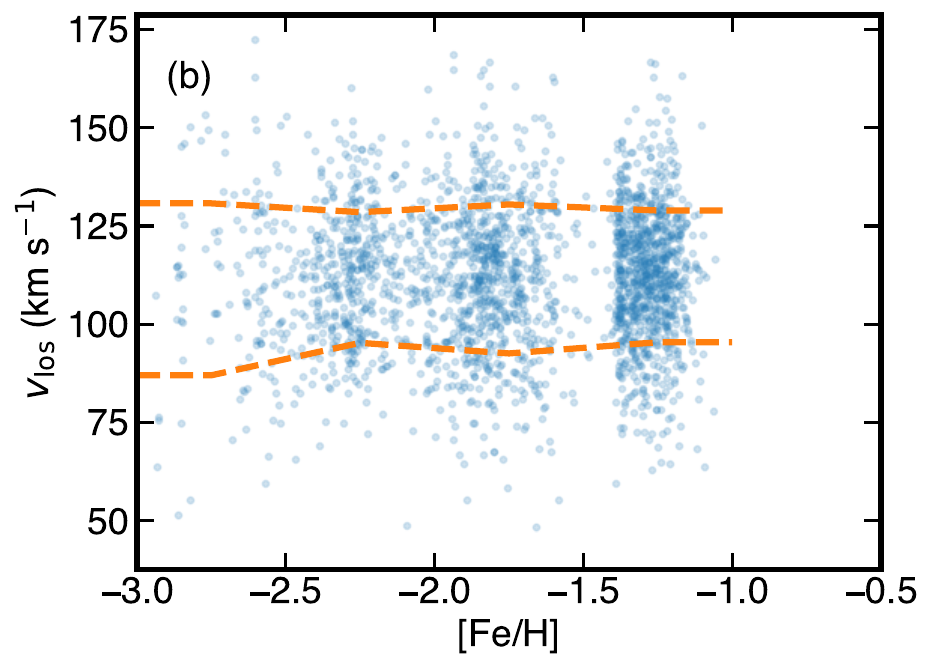}
\caption{Line-of-sight velocities ($v_{\rm{los}}$) as a function of [Fe/H] in (a) Halo 12 and (b) Halo 40. The simulated data do not include simulated observational errors. The orange-dashed line shows the standard deviation of $v_{\rm{los}}$ as a function of [Fe/H].\label{fig:los}}
\end{figure}

The kinematics of stars also differ among stars with different metallicities. Figures \ref{fig:los} (a) and (b) show the line-of-sight velocities ($v_{\rm{los}}$) as a function of [Fe/H]\@. We computed $v_{\rm{los}}$ assuming that Halos 12 and 40 are located in the equatorial coordinates of Fornax and Sculptor \citep{Hayashi2020}, respectively, i.e., we observed Halos 12 and 40 respectively located in the positions of Fornax and Sculptor dSphs from the position of the Sun in the Milky Way. The dispersion of $v_{\rm{los}}$ for [Fe/H] $\leq -1.5$ is 19.3 km s$^{-1}$ in Halo 12 and 19.2 km s$^{-1}$ in Halo 40. On the other hand, stars with [Fe/H] $> -1.5$ have smaller dispersion: 15.0 km s$^{-1}$ (Halo 12) and 16.8 km s$^{-1}$ (Halo 40). These results confirm the existence of kinematical distinct populations in satellites \citep[e.g.,][]{Tolstoy2004, Battaglia2006}.

\section{Discussion} \label{sec:discussion}

\subsection{Chemo-dynamical Evolution of Satellites} \label{sec:chemo-dynamics}

Here, we discuss the chemo-dynamical evolution of the MW's satellites by comparing simulations and observations. The relationship between orbits and SFHs has been argued to explain the variety of observed SFHs seen in MW's satellites. \citet{Miyoshi2020} computed the orbital motions of MW's satellites, including Fornax, Leo I, Sculptor, and Draco, with a time-varying gravitational potential based on the \textit{Gaia} Data Release 2 \citep{Gaia2018} proper motions, and compared them with SFHs. They found that the infall times of classical dSphs coincide well with the peak of the star-formation regions (SFRs), while UFDs had already been quenched before the infall times. 

Simulated satellites have some similarities to galaxies analyzed by \citet{Miyoshi2020}. Halo 12 is similar to the Fornax dSph in terms of its stellar mass and SFH\@. Both galaxies have intermediate age (4--8 Gyr) and old ($>$ 10 Gyr) stellar populations. The orbit of Halo 12 is similar to that of Leo I. Both Halo 12 and Leo I experienced one pericenter passage throughout their orbits. Stellar mass, orbits, and SFHs are similar between Halo 40 and the Sculptor dSph. These galaxies formed most stars prior to their infall. Halo 150 is similar to the Draco dSph regarding stellar mass, orbits, and SFHs. These galaxies also comprise old ($>$ 10 Gyr) stellar populations. These results suggest that star formation in intermediate-age  and old stars in these galaxies was regulated by SN feedback and gas inflow, as we have argued in Section \ref{sec:sfh}.

The major difference between our simulation and the MW's satellites is the star formation after infall. Our simulation does not exhibit enhancement of  the SFR at the time of the infall, which has been observed by \citet{Miyoshi2020}. \citet{DiCintio2021} \citet{} showed that galaxies should satisfy two conditions to enhance the star formation after infall: (1) galaxies must have cold gas with at least 10$^{-2}$ times the virial mass of the halo at the time of the infall and (2) the pericentric distance should be larger than 10 kpc. None of the galaxies analyzed in this study satisfy these conditions.

The strength and treatment of SN feedback highly affect the SFHs and gas outflow of simulated dwarf galaxies. Since 
galaxy formation simulations cannot resolve the evolution of SN remnants, we need to rely on subgrid feedback models \citep[e.g.,][]{Naab2017, Hopkins2018}. \citet{Revaz2012} performed isolated dwarf galaxy simulations with different strengths of SN feedback. Their simulations showed that the star formation lasted $< 1$ Gyr in their strongest feedback case, while stars were continuously formed over 14 Gyr if they adopted a level of feedback 100 times less than the strongest one \citep[also see][]{Hazenfratz2024}. \citet{Xu2022} suggested that the mass-loading factor (the ratio of outflow rate and star formation rate) in dwarf galaxies ($M_{*}\sim 10^4$--10$^7M_{\sun}$) observed in extremely metal-poor representatives explored by the Subaru survey project \citep[e.g.,][]{Kojima2020, Matsumoto2022, Isobe2023, Nishigaki2023, Xu2024} were $\sim$10 to 100 times lower than those predicted in galaxy formation simulations. These results highlight the importance of studying the effects of feedback on the SFHs of dwarf galaxies.

MDFs reflect the SFHs and gas infall/outflow of dwarf galaxies. \citet{Kirby2011b} showed that Fornax dSph has a narrow MDF with $\sigma$ = 0.36\,dex. The Leo I dSph also exhibits a similar MDF. Their chemical evolution model suggested that these galaxies experienced gas infall to shape the narrow MDF. Halo 12 also exhibits a narrow MDF ($\sigma$ = 0.20\,dex) for stars with [Fe/H] $>-1.5$. As described in Section \ref{fig:sfh}, these stars are formed by gas infall. These results suggest that gas infall plays an important role in the chemical evolution of the Fornax and Leo I dSphs.

The Sculptor dSph has a broader MDF ($\sigma$ = 0.46\,dex) than those of the Fornax and Leo I dSphs \citep{Kirby2013}. \citet{Kirby2011b} found that none of their chemical evolution models reproduce Sculptor's MDF. This problem is resolved if they alter the SFH of the chemical evolution model to a more appropriate choice of parameters for SNe Ia and the SFH \citep{Kirby2011a, Homma2015}. \citet{Homma2015} interpreted Sculptor's SFH derived by \citet{deBoer2012a} with a chemical evolution model similar to that of \citet{Kirby2011b}. They found that dSphs with a larger fraction of stars formed in the early phase have a more elongated low-metallicity tail of the MDF.

Halo 40 in our simulation also exhibits a broad MDF ($\sigma$ = 0.46\,dex) similar to Sculptor's MDF. This broad MDF is formed by episodic star formation (Figure \ref{fig:sfh} (b)), rather than the continuous SFH assumed in the one-zone chemical evolution models \citep{Kirby2011a, Homma2015}. From inspection of Figure \ref{fig:mdf} (b), there are at least three distinct peaks in Halo 40's MDF formed by episodic star formation. If this is the case, upcoming spectroscopic surveys of dSphs could confirm whether or not the Sculptor dSph has an episodic SFH (see Section \ref{sec:future}).

\subsection{Prospects for Future Surveys} \label{sec:future}

Identifying whether the MW's satellites have episodic star formation is critical to understanding the effects of SN feedback on their chemo-dynamical evolution and the nature of dark matter \citep[e.g.,][]{Aparicio2001,Bettinelli2019,Rusakov2021}. \citet{Pontzen2012} showed that large-scale bulk motion of gas caused by episodic star formation transforms the cusped density profile of dark matter to a cored one \citep[also see][]{Mashchenko2008, Wheeler2019}. The dependence of SFHs on dark matter profiles in observed satellites is not well understood \citep[e.g.,][]{Hayashi2020, Hayashi2023}. We need additional indicators to identify episodic star formation. As we have found in Figure \ref{fig:alpha}, the episodic star formation creates groups of stars with similar [$\alpha$/Fe] and [Fe/H]\@. We need to search for this feature with observations.

Upcoming wide-field spectroscopic surveys will be able to measure chemical abundances for a sufficiently large number of stars to detect signatures of episodic SFH from chemical abundances \citep[e.g.,][]{Takada2014, Cooper2023}. For example, Subaru PFS will measure Fe and $\alpha$-element abundances for 14,000 and 6,900 stars in Fornax and Sculptor, respectively. In this subsection, we discuss how the simulated [$\alpha$/Fe] vs. [Fe/H] distribution (Figure \ref{fig:alpha}) can be observed by Subaru PFS. 

\begin{figure}[ht!]
\epsscale{1.2}
\plotone{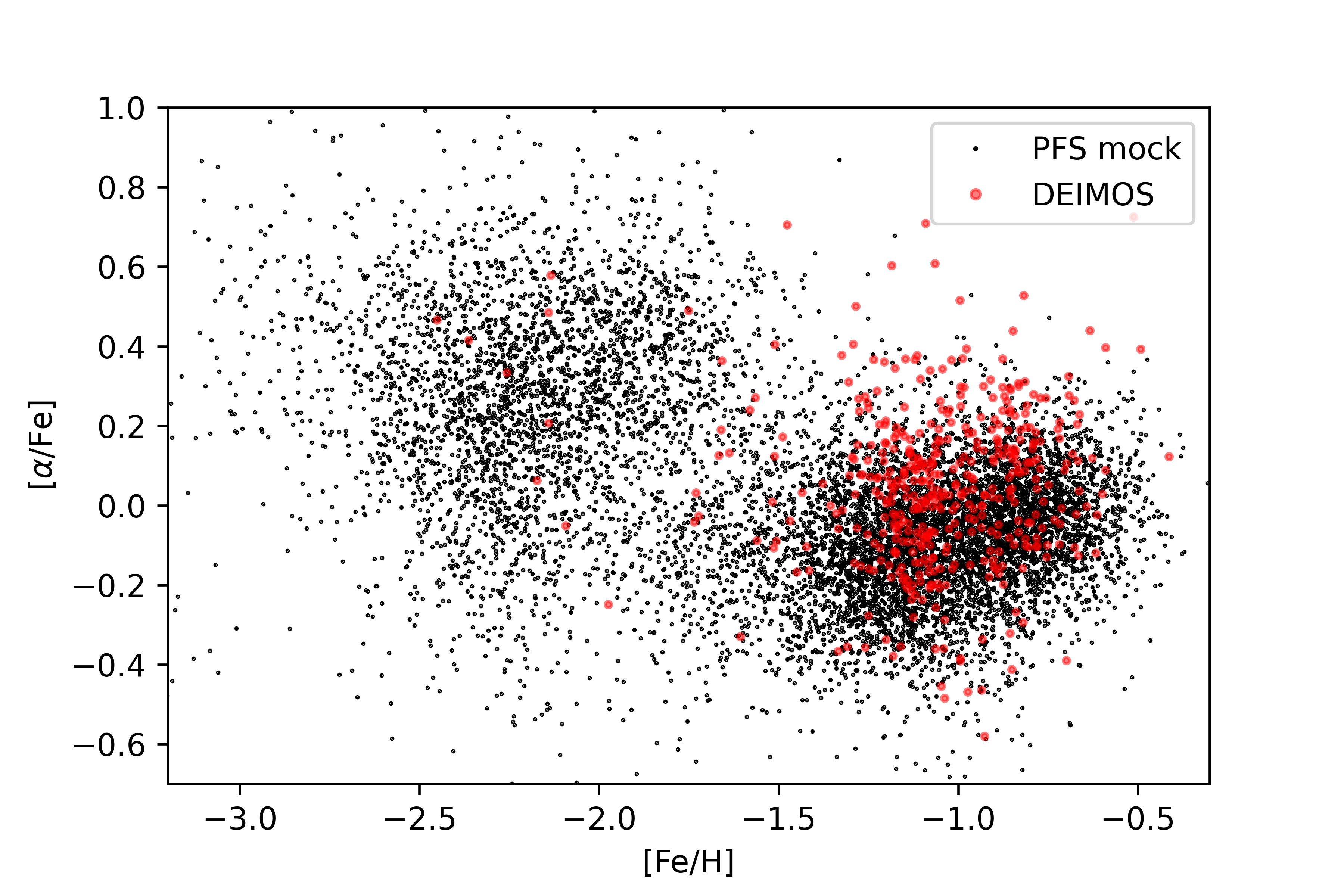}
\plotone{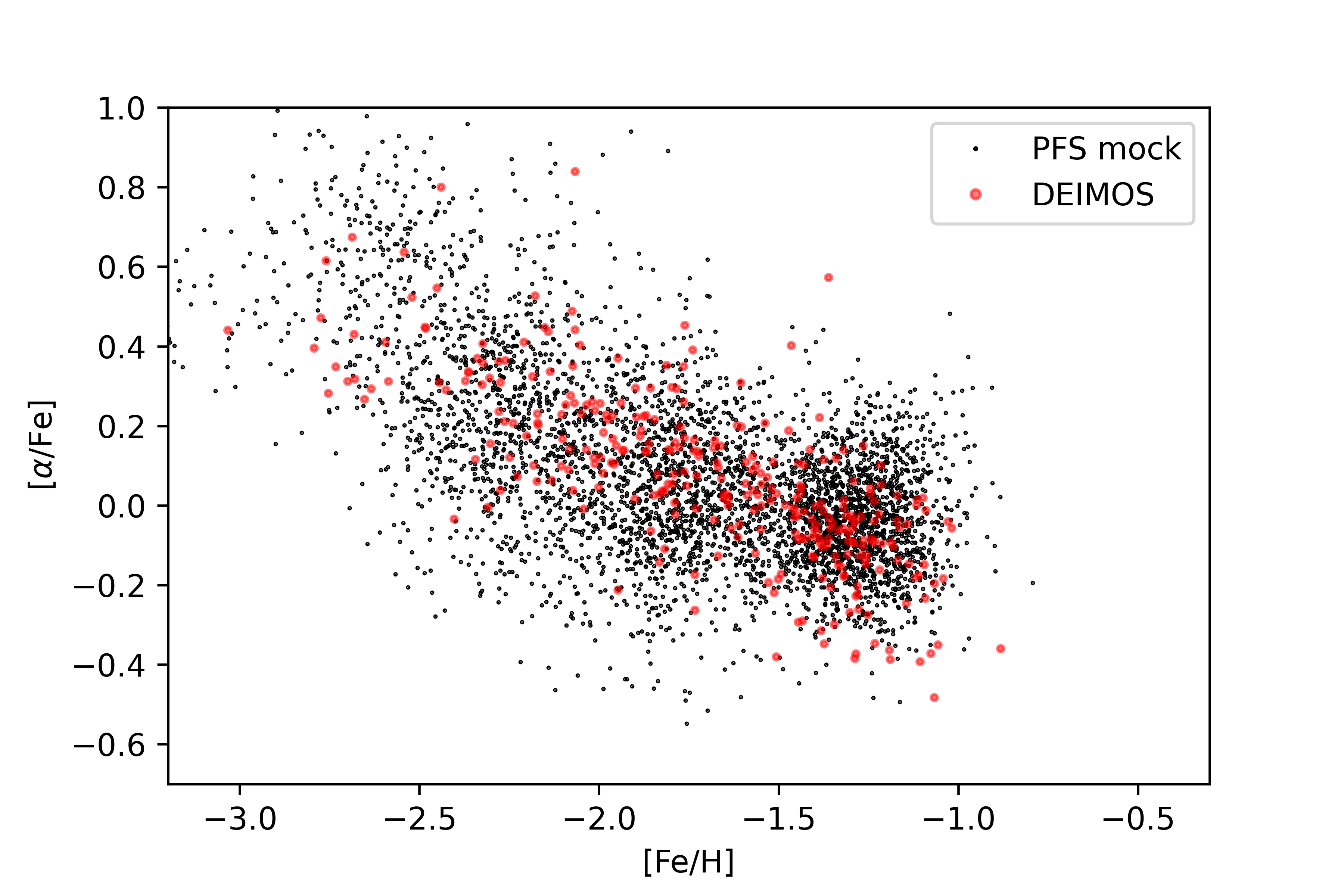}
\caption{Subaru PFS mock observations (black dots) of [$\alpha$/Fe] vs. [Fe/H] for Halos 12 (top panel) and 40 (bottom panel). Red symbols are the abundances for Fornax  (top panel) and Sculptor (bottom panel) observed with Keck/DEIMOS \citep{Kirby2011a}.\label{fig:obs}}
\end{figure}

Figure \ref{fig:obs} shows Subaru PFS mock observations of [$\alpha$/Fe] vs. [Fe/H] for Halos 12 and 40. Procedures for the mock observations are described in Section \ref{sec:mock}. Typical observational uncertainties added to the simulated data are $\sigma\approx$ 0.13\,dex and 0.14\,dex for the [$\alpha$/Fe] and [Fe/H] ratios, respectively. Compared to Figure \ref{fig:alpha}, the scatter in the [$\alpha$/Fe] ratios have been increased. Nevertheless, we can still identify groups of stars having similar [$\alpha$/Fe] and [Fe/H] associated with episodic star formation.

The top panel of Figure \ref{fig:obs} compares mock observed abundances of Halo 12 and the Fornax dSph. With Keck/DEIMOS, \citet{Kirby2011a} found scatter in [$\alpha$/Fe] ratios and a lack of correlation with [Fe/H] in Fornax. Their results suggested that such scatter could arise from bursty star formation or inhomogeneity of the ISM\@. Mock observed [$\alpha$/Fe] ratios in Halo 12 also exhibit scatter for stars with [Fe/H] $>-1.5$. Due to the observed uncertainties, detailed structures of [$\alpha$/Fe] ratios seen in Figure \ref{fig:alpha} (a) cannot be observed, and these structures are observed as scatter. As we have argued in Section \ref{sec:chem}, the scatter of [$\alpha$/Fe] ratios likely come from the enhanced contribution of CCSNe, due to bursty star formation and inhomogeneous chemical abundances in the ISM\@. This result is consistent with the suggestion by \citet{Kirby2011a}.

Stars with [Fe/H] $< -1.5$ in Figure \ref{fig:obs} (top) highlight the importance of observing the Fornax dSph with a wide-field multiplexed spectrograph. In Figure \ref{fig:mdf} (a), we have shown that most stars with [Fe/H] $< -1.5$ are located outside of $r_{\rm{h}}$. Even after applying observed uncertainties, we can still see the decreasing trend of [$\alpha$/Fe] as a function of [Fe/H] and scatter associated with the peaks of episodic star formation. Since the current sample \citep{Kirby2011b} is limited to the center of the Fornax dSph ($\lesssim$ 400 pc), we cannot constrain the chemical evolution in the outskirts of this galaxy. We will be able to investigate the most metal-poor tail of the MDF and [$\alpha$/Fe] ratios by obtaining spectroscopy out to the tidal radius \citep[2,078 pc;][]{Irwin1995} of the Fornax dSph.

There are limitations on the ability of medium-resolution spectroscopy to identify dwarf galaxies accreted to the Fornax dSph with [$\alpha$/Fe] ratios. In Figure \ref{fig:alpha}, we find a low-$\alpha$ ($-0.3<$ [$\alpha$/Fe] $<-0.1$) and very metal-poor ($-2.5<$ [Fe/H] $<-2.2$) component, which is from an accreted dwarf galaxy. However, the distinction of this component is unclear, due to the observed uncertainties in Figure \ref{fig:obs} (top). This result suggests that measuring velocity distribution (Figure \ref{fig:los}) and high-resolution spectroscopy for chemical abundances of stars on the outskirts is necessary to distinguish accreted components. For example, most stars with [Fe/H] $\leq-2.5$ in Halo 12 come from accreted dwarf galaxies. Their line-of-sight velocity dispersion is 22.3 km s$^{-1}$, while that of stars with [Fe/H] $>-2.5$ shows 16.7 km s$^{-1}$ (Figure \ref{fig:los}). These differences in velocity dispersion could be measured in future surveys.

The bottom panel of Figure \ref{fig:obs} compares mock observed [$\alpha$/Fe], as a function of 
[Fe/H], in Halo 40 and the Sculptor dSph. In the mock observation, the groups of stars with similar [$\alpha$/Fe] and [Fe/H] formed in episodic star formation. For [Fe/H] $< -2.0$, these groups are typically separated with 0.5 and 0.4\,dex in [Fe/H] and [$\alpha$/Fe], respectively. However, the number of stars (375) observed in Keck/DEIMOS \citep{Kirby2011b} is insufficient to identify such groups of stars. With Subaru PFS, we expect to measure [$\alpha$/Fe] and [Fe/H] for 6,900 stars in the Sculptor dSph. As shown in this mock observation, in the planned survey we will confirm whether there is episodic star formation occurring every few hundred Myr by identifying chemical clumps.

In this subsection, we have shown that [$\alpha$/Fe] vs. [Fe/H] measured by medium-resolution spectroscopy for $\gtrsim$ 1,000 stars can confirm signatures of episodic star formation in the Fornax and Sculptor dSphs. Thanks to our high-resolution cosmological zoom-in simulation, we can discuss the detailed chemo-dynamical structures of satellite galaxies with $\gtrsim$ 10$^6$ $M_{\sun}$. However, due to the resolution limit, we cannot constrain the SFHs and chemical abundances of galaxies with $\lesssim$ 10$^5$ $M_{\sun}$. The SFHs of poorly resolved galaxies tend to be more bursty, because there are too many synchronized SNe from a star particle \citep[e.g.,][]{Hopkins2018b,Garrison-Kimmel2019}. \citet{Hopkins2018b} showed that simulated galaxies should have $> 100$ star particles to result in a convergence of SFHs. This result means that simulations of MW-like galaxies with a mass resolution of $\sim10\,M_{\sun}$ is required to resolve SFHs of the smallest satellites ($\lesssim10^3\,M_{\sun}$). Such simulations could be achieved by resolving the computational scaling issue using deep learning \citep{Hirashima2023}. We expect that a comparison with upcoming wide-field spectroscopic surveys and high-resolution cosmological simulations will improve our capability to reconstruct the chemo-dynamical evolution of satellites from chemical abundances.
\section{Conclusions} \label{sec:conclusions}
In this study we performed a high-resolution cosmological zoom-in simulation of a MW-like galaxy. With this simulation, we find that the SFHs, MDFs, and [$\alpha$/Fe] ratios of three simulated satellite galaxies are similar to the MW's satellites (Fornax, Sculptor, and Draco dSphs). We also performed a mock observation of medium-resolution spectra using the Subaru PFS spectral synthesis pipeline.

In our simulation, we find that star formation in most simulated satellites is quenched before their infall to the host (Figure \ref{fig:sfh}). Star formation episodes in simulated satellites are separated by a few hundred Myr. Such episodic star formation is regulated by SN feedback. For the Fornax-like galaxy (Halo 12), gas infall induces additional star formation at $\approx$ 6--10 Gyr from the beginning of the simulation.

Simulated MDFs reflect SFHs and gas infall/outflow. The narrow MDF for [Fe/H] $>-1.5$ in Halo 12 is formed in the additional star formation due to gas infall (Figure \ref{fig:mdf} (a)). This feature is similar to the Fornax dSph. In contrast, the Sculptor-mass galaxy (Halo 40) exhibits a broad MDF (Figure \ref{fig:mdf} (b)). This MDF has at least three distinct peaks formed by episodic star formation.

The [$\alpha$/Fe] ratios, as a function of [Fe/H], reflect the ages of stars (Figure \ref{fig:alpha}). The oldest stars have [$\alpha$/Fe] $\gtrsim +0.4$. Subsequent enrichment by SNe Ia decreases the [$\alpha$/Fe] ratios. Stars with similar ages formed in episodic star formation events 
comprise groups with similar [$\alpha$/Fe] and [Fe/H]\@. The bursty star formation and inhomogeneity of the ISM form scattered [$\alpha$/Fe] ratios at [Fe/H] $> -1.5$.

Our mock observations find that the groups of stars with similar [$\alpha$/Fe] and [Fe/H] formed by episodic star formation can be identified by upcoming multiplexed medium-resolution spectra surveys (Figure \ref{fig:obs}). We can test whether satellites have episodic star formation with [$\alpha$/Fe] ratios measured with medium-resolution spectra for $\gtrsim$ 1,000 stars. We also find that metal-poor stellar populations can be found in the outskirts of the galaxy. These results indicate that comparison with upcoming spectroscopic surveys and high-resolution cosmological simulations will greatly improve our understanding of the chemo-dynamical evolution of satellite galaxies.

\begin{acknowledgments}
We are grateful to an anonymous referee for carefully reviewing the manuscript. This work was supported in part by JSPS KAKENHI Grant Numbers JP22KJ0157, JP21H04499, JP21K03614, JP22H01259, JP24K00669, JP20H01895, JP21K13909, JP23H04009, JP22K03688, JP24K07095, JP20H05855, MEXT as ``Program for Promoting Researches on the Supercomputer Fugaku" (Structure and Evolution of the Universe Unraveled by Fusion of Simulation and AI; Grant Number JPMXP1020230406), JICFuS, grants PHY 14-30152; Physics Frontier Center/JINA Center for the Evolution of the Elements (JINA-CEE), and OISE-1927130: The International Research Network for Nuclear Astrophysics (IReNA), awarded by the US National Science Foundation.  E.N.K.\ acknowledges support from NSF CAREER grant AST-2233781. Numerical computations and analysis were carried out on Cray XC50 and computers at the Center for Computational Astrophysics, the National Astronomical Observatory of Japan and the  Yukawa Institute Computer Facility. This research made use of NASA's Astrophysics Data System.
\end{acknowledgments}

\software{AHF \citep{Gill2004, Knollmann2009},
astropy \citep{Astropy2013,Astropy2018}, 
CELib \citep{Saitoh2017},
Cloudy \citep{Ferland2013},
MUSIC \citep{HahnAbel2011}}
          
\begin{figure}[ht!]
\epsscale{1.15}
\plottwo{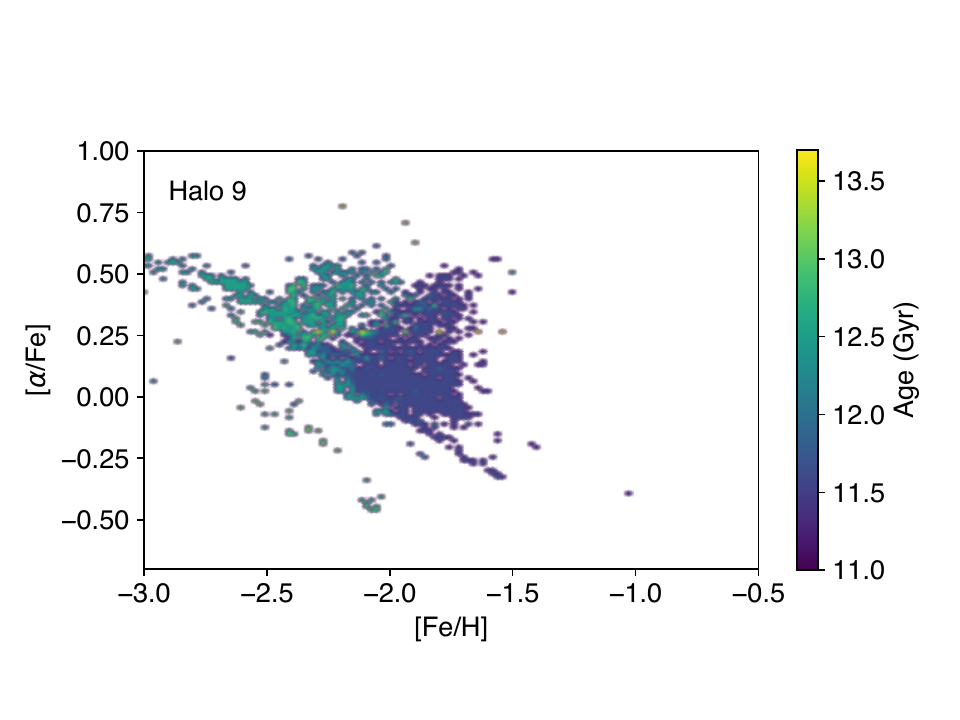}{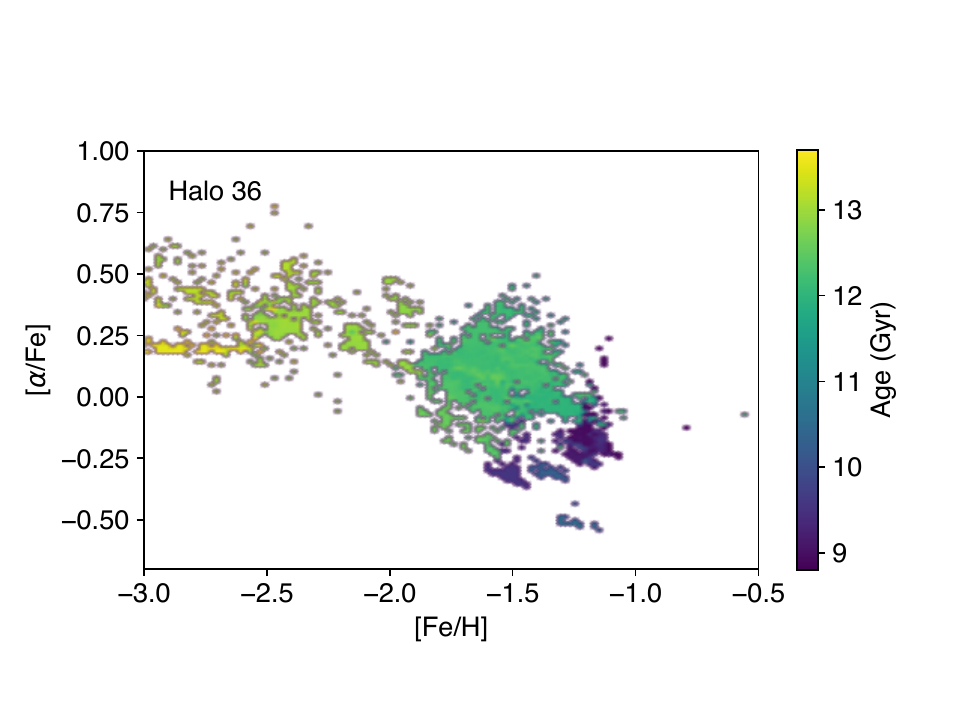}
\caption{Same as Figure \ref{fig:alpha}, but for Halos 9 (left) and 36 (right).  
The color bars indicate the ages of the stars.\label{fig:alpha_all}}
\end{figure}

\appendix
Figure \ref{fig:alpha_all} shows [$\alpha$/Fe], as a function of [Fe/H], for Halos 9 and 36, which have sufficient data to plot. Halo 9 has two major star formation events in 2 Gyr. The first burst forms stars with [Fe/H] $\lesssim -2.0$, and the second burst produces stars with [Fe/H] $\gtrsim -2.0$. As a result of Fe enrichment by SNe Ia, the [$\alpha$/Fe] ratios decrease toward higher metallicity. The inhomogeneity of the spatial metallicity distribution of the ISM due to CCSNe produces scatter of the [$\alpha$/Fe] ratios. On the other hand, Halo 36 has a more extended SFH (Figure \ref{fig:csfh}). The first star formation event creates stars with constant [$\alpha$/Fe] $\approx$ +0.3 due to the contribution from CCSNe. The two subsequent star formation events are influnced by SNe Ia. There is a decreasing trend of [$\alpha$/Fe] ratios toward higher [Fe/H]\@. Similar to Halos 12 and 40, both galaxies have stars with different [$\alpha$/Fe] and [Fe/H], depending on their ages.

\bibliography{references}{}
\bibliographystyle{aasjournal}
\end{document}